\documentclass[notoc]{JHEP} 
\usepackage{amsmath}
\usepackage{epsfig}
\usepackage{amssymb,amsfonts}

\def\be{\begin{equation}}
\def\ee{\end{equation}}
\def\ba{\begin{eqnarray}}
\def\ea{\end{eqnarray}}

\def\tr{\,{\rm tr}\, }

\def\t{\tau}
\def\e{\,{\rm e}}

\newcommand{\ie}{i.e.\ }
\newcommand{\A}{{\cal A}}

\newcommand{\bR}{{\bf R}}

\def\arcsinh{\,{\rm arcsinh}\, }

\def\sqr#1#2{{
\vcenter{\vbox{\hrule height.#2pt
\hbox{\vrule width.#2pt height#1pt \kern#1pt
\vrule width.#2pt}
\hrule height.#2pt}}}}
\boldmath
\title{Anti-de Sitter D-branes}
\unboldmath
\author{C. Bachas\\
Laboratoire de Physique Th{\'e}orique
de l'Ecole Normale Sup{\'e}rieure\thanks{Unit{\'e} mixte  du
CNRS et de l'Ecole Normale Sup{\'e}rieure,  
UMR 8549.}\\ 
24  rue Lhomond, 75231 Paris Cedex 05, France\\
\email{bachas@physique.ens.fr}}

\author{M. Petropoulos\\
Centre de Physique Th{\'e}orique, 
Ecole Polytechnique\thanks{Unit{\'e} mixte  du
CNRS et de  l'Ecole Polytechnique,  
UMR 7644.}\\
91128 Palaiseau Cedex, France\\
\email{marios@cpht.polytechnique.fr}} 

\abstract{We study  D-branes of the ${SL(2,\bR)}$ WZW model 
and of its discrete orbifolds. Gluing the currents by group
automorphisms leads to three types of D-branes: 
two-dimensional  hyperbolic planes ($H_2$),
de Sitter branes (dS$_2$), or  anti-de Sitter branes (AdS$_2$).
We explain  that the dS$_2$ branes are unphysical, because the electric
field on their world-volume is  supercritical. By combining
${SL(2,\bR)}$ and $SU(2)$, we  exhibit
a class of supersymmetric 
AdS$_2\times S^2$ three-brane world-volumes, and consider  their possible
embeddings in the near-horizon region  of
stringy black holes.
We point out the  intriguing difference  between the induced 
and the effective  geometries of these D-branes, and speculate on its
possible significance.
\hskip 10cm
{\sl Dedicated to the memory of
Lochlainn O'Raifeartaigh, who among his many important contributions was
also the first to study string theory on $SL(2,\bR)$.}}

\preprint{LPTENS-0047  
\\CPTH-S046.0400
\\hep-th/0012234}

\begin{document}
\section{Introduction and summary of  results}

The purpose of the present work is to study
D-branes of  the  $SL(2,\bR)$ Wess--Zumino--Witten
model \cite{su1,su11} and of  its  discrete
orbifolds. The corresponding target-space geometries  are
three-dimensional
anti de Sitter (AdS$_3$) with a Neveu--Schwarz three-form flux,
and the three-dimensional anti-de Sitter black holes \cite{btz,bhtz}.
These can be embedded into  exact supersymmetric
backgrounds of string theory \cite{abs,hor}
which arise, in particular, as near-horizon geometries
of stringy black holes \cite{lostro,cvts,ts,hyun,bps,coto,ss,mastro}.
They provide, furthermore, a unique setting in
which to analyse the AdS/CFT correspondence \cite{Maldacena:1998re,adsrev}
beyond the supergravity approximation. The study of D-branes  in these
backgrounds is thus a problem of significant interest.

The WZW  D-branes for compact Lie groups  
have been by now rather  well understood 
\cite{klim,as,fffs,sta,ars,bds,paw,ars2} 
both from  the  conformal field theory (CFT)  
and from the geometric, target-space viewpoint\footnote{
See also  [25--41] for  related works.}.
In the CFT, the symmetric D-branes  are
 described by generalized  Cardy boundary states, 
which are  linear superpositions of Ishibashi
or ``character" states \cite{ca,bppz}.
The  latter couple to closed strings in
a particular representation of the current algebra, while the former have
integer  open-string multiplicities and can be identified with
semi-classical branes. 
From the geometric point of view, these   branes
wrap  special (twined)  
conjugacy classes of the  group manifold \cite{as,fffs,sta}, a fact 
 at first sight paradoxical   because conjugacy classes are
not minimal-area surfaces, and they can contract to a point.
The D-branes carry, however, a 
 quantized world-volume flux whose coupling  with
the background Neveu--Schwarz antisymmetric field
prevents them from shrinking to zero size \cite{bds,paw}.

  For the non-compact group  $SL(2,\bR)$, 
the analogous story has not been  unravelled. Some  results
on symmetric D-branes have been derived in \cite{sta1,sta2},
 but even the
perturbative string theory is not yet fully  understood,
despite significant recent progress \cite{mo,mo1} (see
also \cite{te,gks,bort,ks,long,kas}).
One of our  motivations in  the present work was to 
 gain  more insight into  this important   CFT  
through the semi-classical analysis of D-branes.
In fact, as has been  shown in reference \cite{bds}, 
the semi-classical analysis for compact groups
 gives {\it exact}  results for such CFT  data,   as 
conformal weights  of vertex operators,  and  elements of  the
modular-transformation matrix $S$.  We believe that this is due
to some underlying supersymmetry, and we will therefore discuss
in  detail  supersymmetric settings in which the WZW  D-branes for
both  $SL(2,\bR)$ and $SU(2)$
can be embedded.  Our  analysis 
 will reveal  subtle features that were not present
 in the  case of compact groups: unphysical brane trajectories, 
 quantization conditions that are higher-order in the string coupling,
divergent energies etc. How  these arise from the CFT
is an  interesting question,  to which we hope to return in the near
future.

  One other interesting aspect  of  WZW D-branes
has to do with their  effective world-volume theory. 
 For the spherical D2-branes of
$SU(2)$ this was shown  \cite{ars,ars2}    to be 
a non-commutative field theory on
 the fuzzy two-sphere \cite{ma}.
Within the
semi-classical analysis, the result follows at least qualitatively
from the observation that the antisymmetric-tensor
  flux through the two-sphere
can be made much larger
than the  induced  metric\footnote{This came up in a discussion with
Steve Shenker.}.
One has  thus a curved-space analog of the  decoupling limit
considered in  \cite{cds,dh,sw1}. For the AdS$_2$ D-branes of
$SL(2,\bR)$, which we will exhibit in the present work, 
 the situation turns out  again to be subtler.
The AdS$_2$  branes are the world-volumes of
D-strings carrying  an electric field that can, if desired,
be tuned to its critical limit.  This provides a curved-space
analog of the non-commutative open-string theories,  whose existence
was  conjectured in references \cite{sst,gmms}.

   The plan of this paper is as follows: in Section 2 we will consider 
symmetric gluing conditions for  $SL(2,\bR)$, and discuss the
geometry of the corresponding D-branes. These were  analysed previously 
by Sonia Stanciu  \cite{sta1}, whose  results we will  extend in several
ways. 
We will show, in particular,
 that in addition to regular conjugacy classes,
which are  two-dimensional
 hyperbolic ($H_2$)  or de Sitter (dS$_2$) spaces, there exist twined
conjugacy classes with two-dimensional anti-de Sitter geometry (AdS$_2$).
These are the only ``physically admissible" world-volumes, as we will
explain in the next  two sections. Specifically, in Section 3 we
will see  that the dS$_2$ world-volumes describe, in cylindrical
coordinates,
 circular D-strings 
 which reach the boundary of AdS$_3$
in a finite time.  We will explain why this is impossible, and
identify the ``pathology" as a 
supercritical world-volume electric field. We will also exhibit 
oscillating trajectories, which break the diagonal $SL(2,\bR)$ invariance 
and provide interesting examples of non-symmetric boundary conditions
in WZW models.  In Section 4 we will analyse similarly the AdS$_2$
world-volumes, which  correspond to static $(1,q)$ strings
stretching across antipodal points on the AdS$_3$ boundary.
We will  show that these AdS$_2$ world-volumes are the only 
static solutions of the Dirac--Born--Infeld equations with Neumann
boundary conditions along the Poincar{\'e} horizon.
 
  In Section 5 we will combine the  $SL(2,\bR)$ and $SU(2)$ branes,
in order  to preserve some space-time supersymmetries. The resulting
AdS$_2$$\times $$S^2$  D3-branes respect, as we will show,
half  of the supersymmetries
of the ambient AdS$_3$$\times $$S^3$ geometry, which  arises in  the
near-horizon region  of the NS5/F1 black
string. In   Section 6 we discuss  these BPS D3-branes from
the  perspective of their world-volume. One 
striking fact is that,  whereas  the induced radius of  AdS$_2$
can be made arbitrarily large, and that of 
 $S^2$  arbitrarily small,  the
 effective open-string  radii stay  equal to each other, and 
to the radius  of the ambient geometry. This is  required
by world-volume  supersymmetry,
and points towards  a more general property of D-branes.
Interestingly,
the effective geometry on the D3-brane 
turns out to be the same as for  an extremal
four-dimensional dyonic black hole.
In Section 7 we will investigate the fate
of these  D-branes in the background of a BTZ black hole,
which is obtained by discrete identifications
of (a part of) AdS$_3$ space-time.
The BTZ geometry arises  in the near-horizon region of 
the four- and  five-dimensional black holes used in the
microscopic entropy
calculation \cite{sv, cm, ss, mastro}.
We will explain why   supersymmetry is always broken
by our D-branes in this case, but   could be  restored 
for more general orbifolds possibly related to 
spinning five-dimensional black holes \cite{cl,mm}.  

We have collected, for
the reader's convenience, the various
coordinate systems for AdS$_3$ in Appendix A. In Appendix B
we reanalyse the AdS$_2$ branes in global 
(rather than Poincar{\'e}) coordinates, which are better
adapted for the study of small fluctuations. Appendix C contains
a brief discussion of the instatonic $H_2$ branes.  


\vskip 0.5cm
\boldmath
\section{(Twined) conjugacy classes of $SL(2,\bR)$}
\unboldmath

All $n$-dimensional  
manifolds of constant curvature, and with either Euclidean or Lorentzian
signature, can be described as pseudospheres
embedded in $n+1$ flat dimensions,
\begin{equation}
X^M X_M \equiv \epsilon_0 \left(X^0\right)^2 
              +            \left(X^1\right)^2 
              +\cdots +    \left(X^{n-1}\right)^2 
              + \epsilon_n \left(X^n\right)^2
              = \epsilon  L^2     .
\label{pseudo}
\end{equation}
Here $\epsilon_0 , \epsilon_n , \epsilon = \pm $ are
signs, and the ambient space has signature 
$(\epsilon_0, + , \ldots, +, \epsilon_n)$. 
Correspondingly, the pseudosphere has $SO(n+1)$, $SO(n,1)$ or
$SO(n-1,2)$ isometry. 
The different possibilities are summarized in the table below.

\vskip 1cm
\begin{table}[htp]
\begin{center}
\begin{tabular}{|c|c|c|}
\hline
&&
\\
$(\epsilon_0 , \epsilon_n , \epsilon)$   &  Space &
Signature
\\
&&
\\
\hline\hline
&&
\\
--  --  --  & AdS$_n$ & Lorentzian
\\
&&
\\
\hline
&&
\\
-- + +  & dS$_n$  & Lorentzian
\\
&&
\\
\hline
&&
\\
-- + --  & $H_n$  & Euclidean
\\
&&
\\
\hline
&&
\\
+ + +  & $S^n$ & Euclidean
\\
&&
\\
\hline
\end{tabular}
\end{center}
\caption{
The four (pseudo)spheres described by Eq. (\ref{pseudo})
and which have at most one time-like coordinate in $n$ dimensions. }
\end{table}

The two missing entries in this table are   $(- - +)$ which gives a pseudosphere
with two time-like coordinates, and
$(+ + -)$ which gives an equation with no solutions. The space $H_n$ is
the hyperbolic space of constant negative curvature,  also
called  Euclidean AdS  or Lobatchevski plane. For the
ordinary, Lorentzian  AdS, 
the time coordinate is the angular coordinate in the $(X^0,X^n)$ plane.
To avoid closed time-like curves  one must  ignore this
periodic 
identification and consider the full
covering space,  obtained  by gluing together an infinite
number of pseudospheres. All of the above  spaces have Ricci
scalar curvature equal to $\epsilon n (n-1)/L^2$. 

We turn  now to  the special case of AdS$_3$,  which is the
(universal cover of the) group manifold 
of the non-compact group $SL(2,\bR)$. 
A general  group element can be parametrized as follows:
\begin{equation}
g  = {1\over L}
\left(
\begin{array}{lll}
X^0+X^1 &\quad & X^2+X^3\cr
X^2-X^3 &\quad & X^0- X^1\cr
\end{array}
\right) ,
\label{param}
\end{equation}
so that Eq. (\ref{pseudo}) is the condition that the determinant
of the matrix be equal to one.
We are interested in the  (twined) conjugacy classes
\begin{equation}
{\cal W}_g^\omega = \left\{ \omega(h)  g h^{-1}, \ 
\forall
h \in SL(2,\bR) \right\} ,
\end{equation} 
where $\omega$ is an
automorphism   and $g$  a fixed element of the group.
The ${\cal W}_g^\omega$ are the  world-volumes of  D-branes
obtained by identifying,  modulo $\omega$,  the left and
right moving $SL(2,\bR)$ currents of the WZW model \cite{as,fffs,sta}. 
Since these boundary conditions preserve a larger symmetry than
Virasoro, the corresponding D-branes are not generic. 
They are however simplest to discuss, and we will focus our attention
on them in what follows.

When $\omega$ is an inner automorphism, $\omega(h) = g_0^{-1} h\,
g_0^{\vphantom 1}$, so that 
${\cal W}_g^\omega$  is  the  (left) group
translation  of some regular conjugacy class of the group
 (the class  of the element
$ g_0 g$). It is thus sufficient to take
$g_0 ={\bf 1}$, and then
consider group transformations of the resulting trajectories\footnote{More 
generally there is  a non-trivial (left and right)
$SL(2,\bR)\times SL(2,\bR)$ action that gives new classical
trajectories from old ones. Since our D-branes  preserve a diagonal
symmetry, only half of this action is non-trivial in our case.}.  
In the parametrization \eqref{param} the regular conjugacy classes are
described  by
\begin{equation}
\tr g = 2X^0/L  = 2 \tilde C 
\label{constraint} 
\end{equation}
for some constant  $\tilde C$.
Plugging this  into equation 
(\ref{pseudo}) one finds: 
\begin{equation}
\left(X^3\right)^2 -  \left(X^1\right)^2 -  \left(X^{2}\right)^2 =
L^2 \left( 1-{\tilde C}^2 \right) .
\label{trds}
\end{equation}
The nature of this world-volume depends on  whether 
$\big\vert \tilde C \big\vert$ is bigger, equal  or smaller than one.
From  Table 1 we  see  that for $\big\vert \tilde C \big\vert < 1 $
the world-volume is a  hyperbolic plane ($H_2$), for $\big\vert \tilde C
\big\vert > 1 $
it is  two-dimensional de Sitter space  (dS$_2$), while in the
special case $\big\vert \tilde C \big\vert =  1 $ it degenerates into
(part
of)
the light cone in three dimensions. 
Strictly speaking this  latter breaks up into three
distinct conjugacy classes (the apex, the future cone 
and the past cone), while
for $\big\vert \tilde C \big\vert < 1 $ 
one finds  also two disjoint hyperbolic planes. 
These D-brane world-volumes have been described  previously by
Jose Figueroa-O'Farrill and Sonia Stanciu
 \cite{sta1,sta2}. Note that the hyperbolic
plane has Euclidean signature and must thus be interpreted as an
instanton. As we will show  in the following section, the dS$_2$
world-volumes are also ``unphysical" because they occur in a region of
supercritical electric field,  where the D-brane action is imaginary.
None of the above world-volumes corresponds, therefore,
to a physically-acceptable D-brane motion. 

\vskip 0.1cm
\begin{table}[htp]
\begin{center}
\begin{tabular}{|c|c|}
\hline
&
\\
 Conjugacy class    &  D-brane 
\\
&
\\
\hline\hline
&
\\
 $-\infty < \tr (\omega_0 g) <\infty$   & AdS$_2$ 
\\
&
\\
\hline
&
\\
$\vert \tr g \vert < 2$   & $H_2$ 
\\
&
\\
\hline
&
\\
$\vert \tr g \vert > 2$ & dS$_2$  
\\
&
\\
\hline
&
\\
 $\vert \tr g \vert =  2$  & light cone  
\\
&
\\
\hline
&
\\
 $g =   {\bf 1}$  & point
\\
&
\\
\hline
\end{tabular}
\end{center}
\caption{The different (twined) conjugacy classes of $SL(2,\bR)$ and
the geometry of the corresponding D-brane world-volumes. All
world-volumes have dimension two, except for the degenerate case $g= 
{\bf 1}$,  which corresponds to a point-like D-instanton.}
\end{table}

Consider next the case  when $\omega$ is
an outer automorphism.
The non-trivial  outer automorphism 
of $SL(2,\bR)$, up to group conjugation,  is the operation that
changes the sign of  $X^1$ and $X^3$, while  leaving  $X^0$ and $X^2$
unchanged. In terms of $2\times 2$ matrices we may write 
\begin{equation}
h \to \omega_0^{-1}  h\,  \omega_0^{\vphantom 1} \  , \ \ {\rm with} \ \ 
\omega_0^{\vphantom 1} = \left( \begin{array}{ll}
0 & 1 \cr
1 & 0 \cr
\end{array}
\right) .
\label{outer}
\end{equation}
Since  $\omega_0$ is not an element of the group (its determinant
is minus one) this  automorphism is indeed not inner. The D-brane
world-volume corresponding to this  gluing condition is the
twined  conjugacy class   
\begin{equation}
\tr (\omega_0 g) = 2X^2/L = 2{C} \, , 
\label{adscon}
\end{equation}
with $ C$ a constant. For any  value of $ C$ the world-volume geometry
is, in this case, two-dimensional 
anti-de Sitter (AdS$_2$):
\begin{equation}
\left(X^0\right)^2 + \left(X^3\right)^2 - \left(X^{1}\right)^2 =  L^2
\left(1+{ C}^2\right). 
\label{trads}
\end{equation}
We will show in Section 4 that the AdS$_2$ geometries 
are physically-acceptable world-volumes of stretched D-strings.
Combining $\omega_0$ with non-trivial  inner
automorphisms  gives other physical trajectories, which are  
group translations of \eqref{trads}. 

Table 2 summarizes for convenience all the (twined) conjugacy classes of
$SL(2,\bR)$. In the CFT, the corresponding
D-branes should be appropriate superpositions of  
Ishibashi boundary
states of the
current algebra. Their algebraic construction is a very  interesting and subtle
problem,  which we will  not address in the present work. We will
  focus instead
on  their  geometric, semi-classical interpretation. 


\vskip 0.5cm
\boldmath
\section{Dynamics of a circular   D-string}
\unboldmath

There are several different coordinate systems for  AdS$_3$
that  will be useful in our analysis -- we have collected them all for 
convenience in Appendix A. The global structure is  easier 
to visualize in the cylindrical coordinates
\begin{equation}
X^0 + i X^3 = L  \cosh\rho \e^{i \tau} \ , \ \ X^1 + i X^2 =
L \sinh\rho \e^{i\phi}\, , 
\label{cyl}
\end{equation}
in which 
the metric and Neveu--Schwarz antisymmetric tensor
backgrounds  of the WZW model read:
\begin{equation}
ds^2 = L^2 \left(-  \cosh^2\! \rho \,
d\tau^2 + d\rho^2  + \sinh^2\! \rho\, d\phi^2 \right) ,
\end{equation}
and 
\begin{equation}
H = dB =  L^2  \sinh(2\rho)\, d\rho\wedge d\phi\wedge d\tau \, .
\end{equation}
The three-form field strength is the volume
form of the manifold,  up to a constant of proportionality that is
fixed by the conformal-invariance conditions.
The coordinate  $\phi$ is  an angular variable, $\rho$ is
a radial variable taking values in $ [0,\infty[$, and  
$\rho\to\infty$ is the
boundary of the anti-de Sitter space. The radius of AdS$_3$ is given by
the  level $\tilde k$ of the associated $SL(2,\bR)$  current algebra. In
the semi-classical, large-radius limit 
$L^2 \simeq \big\vert \tilde k \big\vert \alpha^\prime$. In contrast to
the compact case 
$SU(2)$, the level $k$ need not be here integer,  since the $B$-field
has no Dirac singularity anywhere in the interior of space-time.  

In  cylindrical coordinates
 the d$S_2$ conjugacy class  is given  by
\begin{equation}
X^0/L = \cosh \rho \, \cos \t = \tilde C >1 \, ,
\label{ds2}
\end{equation}
where  $\tilde C$ is constant. This is the world-volume
of a circular string moving in from the boundary of AdS$_3$ to some
minimum radius, $\cosh \rho_{\min} = \tilde C$,  and then out  to the
boundary again. The entire motion occurs over 
a time interval $\Delta \t = \pi$. This
motion is forbidden classically,
because only a massless particle may  reach the boundary of AdS, and 
our  D-strings are a priori massive. 

To see  why a  massive  particle can never reach the boundary of AdS, 
consider its energy as measured by an
observer sitting at the center, 
\begin{equation}
E L = {\partial {\cal L}\over \partial {\dot\rho}} {\dot\rho} - {\cal L} = 
{m L \cosh^2\! \rho \over 
\sqrt{\cosh^2\! \rho-{\dot\rho}^2}}\, . 
\end{equation}
Here $m$ is the mass of the particle, ${\cal L}$ the relativistic
point-particle Lagrangian,   the
dots stand for derivatives with respect to 
 $\tau$, and we have restricted our attention for simplicity to radial
motion.   This energy is bounded from below by
the blue-shifted
mass, $E \ge m  \cosh \rho$, \ie by the mass of the particle when at rest at
radius $\rho$.
Bringing  the particle
to the boundary, where the blue shift 
diverges, would  require an
infinite energy, and is thus not allowed. 

The extension of the  argument to a  $(p,q)$ string requires  some care,
because of the extra coupling to the $B$-field  background. 
A fundamental ``long string", in particular, 
{\it can}  reach  the boundary of AdS$_3$
 despite the infinite blue shift,  and even
though its  length also diverges \cite{mo,long}.
The reason for this is that the infinite tensive energy cancels
against the divergent $B$-field potential.
Such a mechanism does not, however, work for a general  $(p,q)$ string, whose
tension is greater than its  Neveu--Schwarz charge density.
Explicitly, if we denote by $T_{\rm D}$ and $T_{\rm F}$ the 
D-string and fundamental-string
tensions, then
\begin{equation}
T_{(p,q)} =  \sqrt{p^2_{\vphantom D} T_{\rm D}^2 + q^2_{\vphantom D}
T_{\rm F}^2} > \rho_B = qT_{\rm F} \, .
\end{equation}
Thus,  tension 
dominates  near the boundary of AdS, which for all but pure fundamental strings 
($p=0$) is
 an energetically-forbidden region\footnote{ F-strings and $(p,q)$ strings
are  exchanged by  $SL(2,{\bf Z})$ duality, but the symmetry is here
broken by the NS  background.}.

We can make  the argument more precise, by
considering  the  Dirac--Born--Infeld action for  a D-string  (see
for instance \cite{revs})
\begin{equation}
I  = \int d^{2} \! \zeta \, {\cal L} 
\ , \ \ 
{\cal L} =  - T_{\rm D} \sqrt{-\det \left(
\hat g +\hat B + 2\pi \alpha^{\prime} F
\right)} \, . 
\label{IDBI}
\end{equation}
Here  $\hat g$ and $\hat B$ are the pull-backs of the WZW backgrounds, and
$F$ the world-volume electric field. 
We are  neglecting
higher-order curvature  corrections to the DBI action, 
and we have also dropped the Wess--Zumino terms,
because all RR  backgrounds vanish.
We may  choose a gauge in which the NS  background reads:
\begin{equation}
B= L^2  \sinh^2\! \rho \, d\phi\wedge d\t \, .
\label{bf}
\end{equation}
Using $\phi$ and $\tau$ to parametrize the D-string world-volume,
and keeping only the global breathing
mode $\rho(\phi,\tau)= \rho(\tau)$, leads to the following action 
for  a circular D-string: 
\begin{equation}
I = - 2\pi T_{\rm D} 
\int d\tau \sqrt{-\det \hat g  - {\cal F}_{\phi \t}^2 }\, ,
\label{gflag}
\end{equation}
where 
\begin{equation}
\det \hat g  = - L^4  
\sinh^2\! \rho \left(\cosh^2\! \rho - {\dot\rho}^2\right)  , 
\end{equation}
and 
\begin{equation}
{\cal F}_{\phi \t} =  L^2  \sinh^2 \! \rho - 2\pi\alpha^\prime
 {\dot A}_\phi \, .
\end{equation}

The Wilson line $A_\phi$ is a cyclic variable: it is 
periodic with period one
and its conjugate
momentum is a quantized constant of the motion,
\begin{equation}
{1\over 2\pi}\, \Pi_\phi \equiv 
{\partial {\cal L}\over \partial {\dot A}_\phi}=
- {2\pi\alpha^\prime \, T_{\rm D} \, {\cal F}_{\phi \t}\over \sqrt{-
\det \hat g - {\cal F}_{\phi \t}^2 } } = -q \in {\bf Z} \, .
\label{quant}
\end{equation}
The integer  $q$ is the number of oriented fundamental strings
bound to the
D-string \cite{revs}.
The other constant of the motion is the energy, 
 as measured by an observer
sitting at the  center
of AdS$_3$:
\begin{equation}
EL = 2\pi
\left( {\partial {\cal L}\over \partial {\dot\rho}} {\dot\rho} +
 {\partial {\cal L}\over \partial {\dot A}_\phi}{\dot A}_\phi - 
{\cal L}\right) = {2\pi T_{\rm D}} \,
{ L^4 \sinh^2\! \rho \,  \cosh^2\! \rho
- B_{\phi \t}\,  {\cal F}_{\phi \t}\over \sqrt{-\det \hat g
 - {\cal F}_{\phi \t}^2 }}\, . 
 \label{engen}
\end{equation}
Using Eqs. \eqref{bf} and \eqref{quant}, as well as  the derived  relation
\begin{equation}
 {T_{\rm D}\over \sqrt{-\det \hat g  - {\cal F}_{\phi \t}^2 }} 
=  {T_{(1,q)}\over \sqrt{- \det \hat g } }\, ,
\label{derived}
\end{equation}
we can put  the energy in the more suggestive form
\begin{equation}
E  = {M(\rho)  \cosh^2 \! \rho \over 
\sqrt{\cosh^2 \! \rho - {\dot\rho}^2}} - 
{2\pi q L T_{\rm F}} \sinh^2\! \rho\, ,
\label{a}
\end{equation}
where 
\begin{equation}
M(\rho) = 2\pi L T_{(1,q)}  \sinh \rho
\end{equation}
is the (effective) 
 mass of a circular $(1,q)$ string sitting at radius $\rho$
in AdS$_3$ space.

Equation \eqref{a} exhibits the two competing terms of the potential energy:
the blue-shifted   mass,  and the  interaction with the $B$-field
potential.
Because $ T_{(1,q)}  > \vert q\vert T_{\rm F}$, the first term 
diverges faster than the second near the boundary,
which is thus a forbidden region as advertised.
For pure
fundamental strings,
on the other hand, the two divergent terms 
can  cancel  in the
asymptotic region, leading to a finite
energy  $E\ge \pi q L T_{\rm F}$.
This agrees with the CFT analysis 
of \cite{mo,mo1},  where the energy of a long fundamental string  was
found to be $E \ge  \pi \varpi L T_{\rm F}$; $\varpi$ stands here for the
spectral-flow 
integer parameter that turns out to be a string winding number. Notice
that
$q$ must here be positive: the long strings have a preferred  orientation 
in the Wess--Zumino--Witten  background.

\vskip 1.5cm
\FIGURE{\begin{picture}(300,150)(0,0)
\put(50,0){\epsfig{file=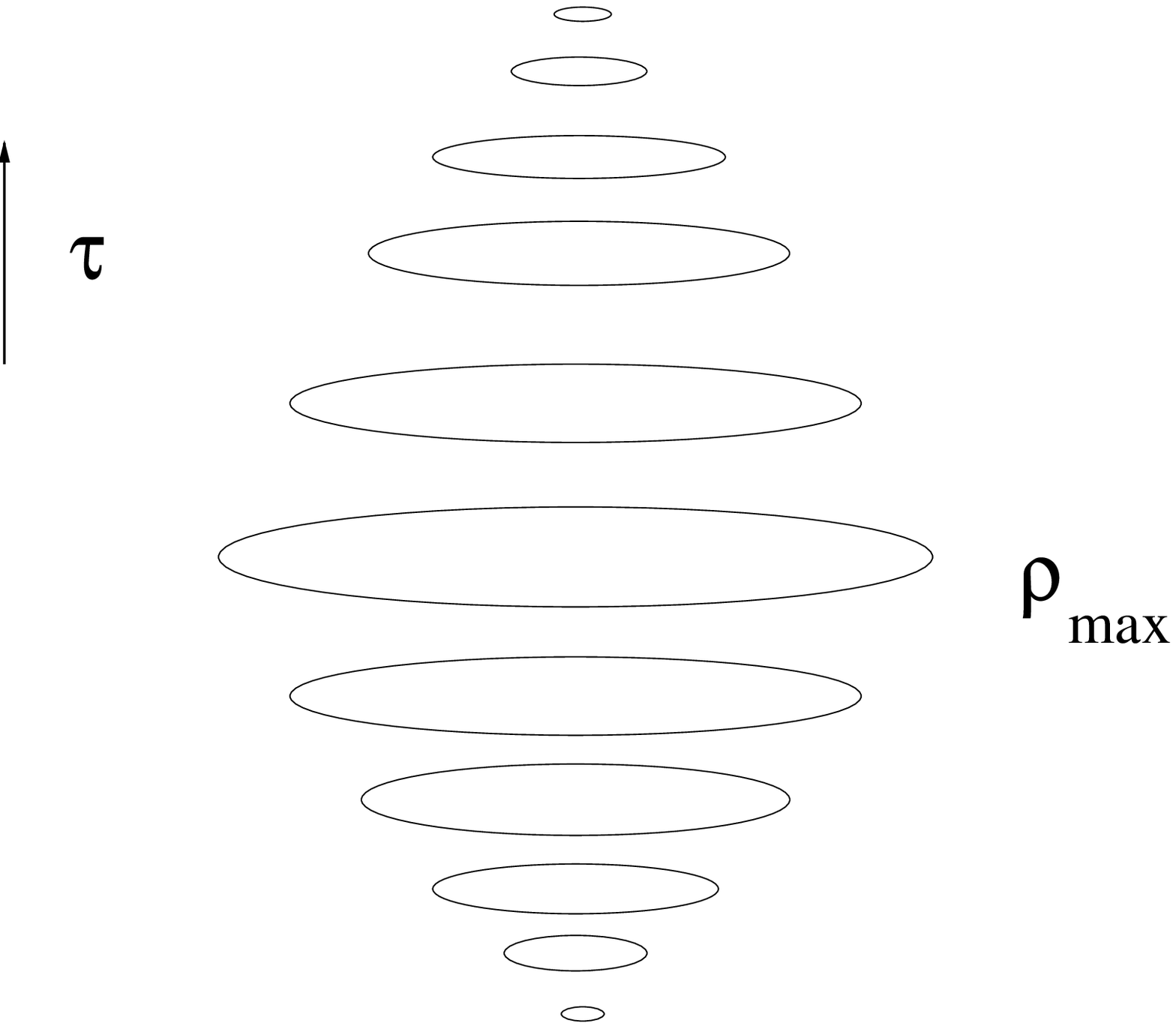,height=6cm}}
\end{picture}
\caption{A circular  $(1,q)$ string performing a  periodic breathing
motion
in AdS$_3$.
\label{radial}}}
\vskip 0.3cm

The radial motions of a  D-string can be obtained easily
from Eq. \eqref{a}, by first 
solving  for $\dot\rho$ as function of $\rho$,  and then integrating
the time.  They  are periodic movements  in which
the string goes out to a maximum radius $\rho_{\rm max}$ before recollapsing
to the center of  AdS$_3$  (see Figure 1). 
For sufficiently  high energy the maximum radius  is 
\begin{equation}
\rho_{\rm max} \approx
 {1\over 2}{\log }\left({2E/\pi L \over   T_{(1,q)}-qT_{\rm F}}\right) .
\end{equation}
This diverges in the (formal) limit $T_{(1,q)}\to qT_{\rm F}$,
consistently with the fact that fundamental strings {\it can}  indeed  reach 
the boundary of AdS$_3$ space. Notice that the above  world-volumes
provide 
an  example  of
non-symmetric D-branes. Left and right currents in the CFT are not
glued by a global automorphism -- if they were we would have obtained
one of the world-volumes in  Table 2. 

There is one final point to address: 
the  dS$_2$ world-volumes  were obtained from   conformally-invariant
gluing conditions, so  they must satisfy  automatically the (target-space) 
Dirac--Born--Infeld
equations. How come then that such motions are not allowed? The answer
is that they are solutions of the DBI equations, indeed, but
in the unphysical region of supercritical electric
field,  ${\cal F}_{\phi \t}^2 > -\det\hat g$.
We can  check this  by
plugging  \eqref{ds2} in the expression for the energy,
which  is  a (vanishing) constant of the motion
only  if  $q$ is allowed to become  imaginary.

An alternative derivation can be given in the 
system of  coordinates
\begin{equation}
X^0 = L \cosh \tilde\psi \ , \ \ 
X^3 = L \sinh \tilde\psi \, \sinh  {\tilde t} \ , \ \ 
X^1+iX^2 = L \sinh \tilde\psi \, \cosh {\tilde t}   \e^{i\phi} \, ,
\label{coords}
\end{equation}
in which the dS$_2$ world-volumes are sections  at
fixed values of  $\tilde\psi$ (see Appendix A). The
string metric and 
Neveu--Schwarz potential  in a convenient 
gauge read:
\begin{equation}
ds^2 = L^2 \left[ d\tilde\psi^2 +
\sinh^2 \! \tilde\psi  \left( -d{\tilde t}^{2} +
\cosh^2\! {\tilde t} \,  d\phi^2 \right) \right]
\end{equation}
and
\begin{equation}
B =  L^2
\left( {\sinh 2\tilde\psi \over 2} -  \tilde\psi \right)
\cosh {\tilde t} \,   d\phi \wedge d{\tilde t} \, .
\label{bf1}
\end{equation}

We use ${\tilde t}$ and $\phi$ as world-volume coordinates, and turn on
a covariantly-constant electric field
\begin{equation}
F = dA =   a L^2  \cosh {\tilde t} \, d\phi \wedge d {\tilde t}\, ,
\end{equation} with $a$ constant. A simple calculation leads to
the following DBI action for a fixed-$\tilde\psi$ trajectory:  
\begin{equation}
I(a,\tilde\psi)=  - 2\pi L^2 \,  T_{\rm D}  \int^{+\infty}_{-\infty}
d{\tilde t} \, \cosh {\tilde t}\,  
\sqrt{\sinh^4 \! \tilde\psi -
\left({\sinh 2\tilde\psi \over 2} -\tilde\psi
+{2\pi\alpha^\prime a} \right)^2 }\, .
\label{acds}
\end{equation}

A dS$_2$ world-volume with a covariantly-constant electric field
will be a solution of the full D-string equations, provided we
extremize the DBI action with respect to $\tilde\psi$. This is
guaranteed by the unbroken $SL(2,\bR)$ symmetry.
Solving the $\tilde\psi$ equation
gives $\tilde\psi = 2\pi\alpha^\prime  a$,  or
$\tilde\psi
=0$. The latter is a  degenerate  solution, while the
former is precisely
our  d$S_2$ world-volume,  with
$\tilde C=\cosh 2\pi\alpha^\prime a $.
One can check that for all $a\not=0$  the expression
inside the square root above is negative,
i.e. the  world-volume field is
supercritical.   This shows that the dS$_2$ solution
is indeed  physically-unacceptable, as advertised.
 
\boldmath
\section{Stretched static D-strings}
\unboldmath

Let us  proceed next  to analyse  the D-branes 
with AdS$_2$ geometry. In cylindrical coordinates, the
defining relation \eqref{adscon}
reads:
\begin{equation}
\sinh \rho \, \sin \phi = {C} \, .
\label{adscon1}
\end{equation}
This is the world-volume of a  static string stretching between two
antipodal points ($\phi=0,\pi$) 
 on the boundary of the ambient AdS$_3$ space-time,
see Figure 2. The minimal radius is 
\begin{equation}
\rho_{\rm min} = \arcsinh  C \, .
\end{equation}
We will see in a moment  that $C= qT_{\rm F}/T_{\rm D}$, where $q$ is the
number of fundamental strings  bound to our static  D-string.

\vskip 1.5cm
\FIGURE{\begin{picture}(300,150)(0,0)
\put(50,0){\epsfig{file=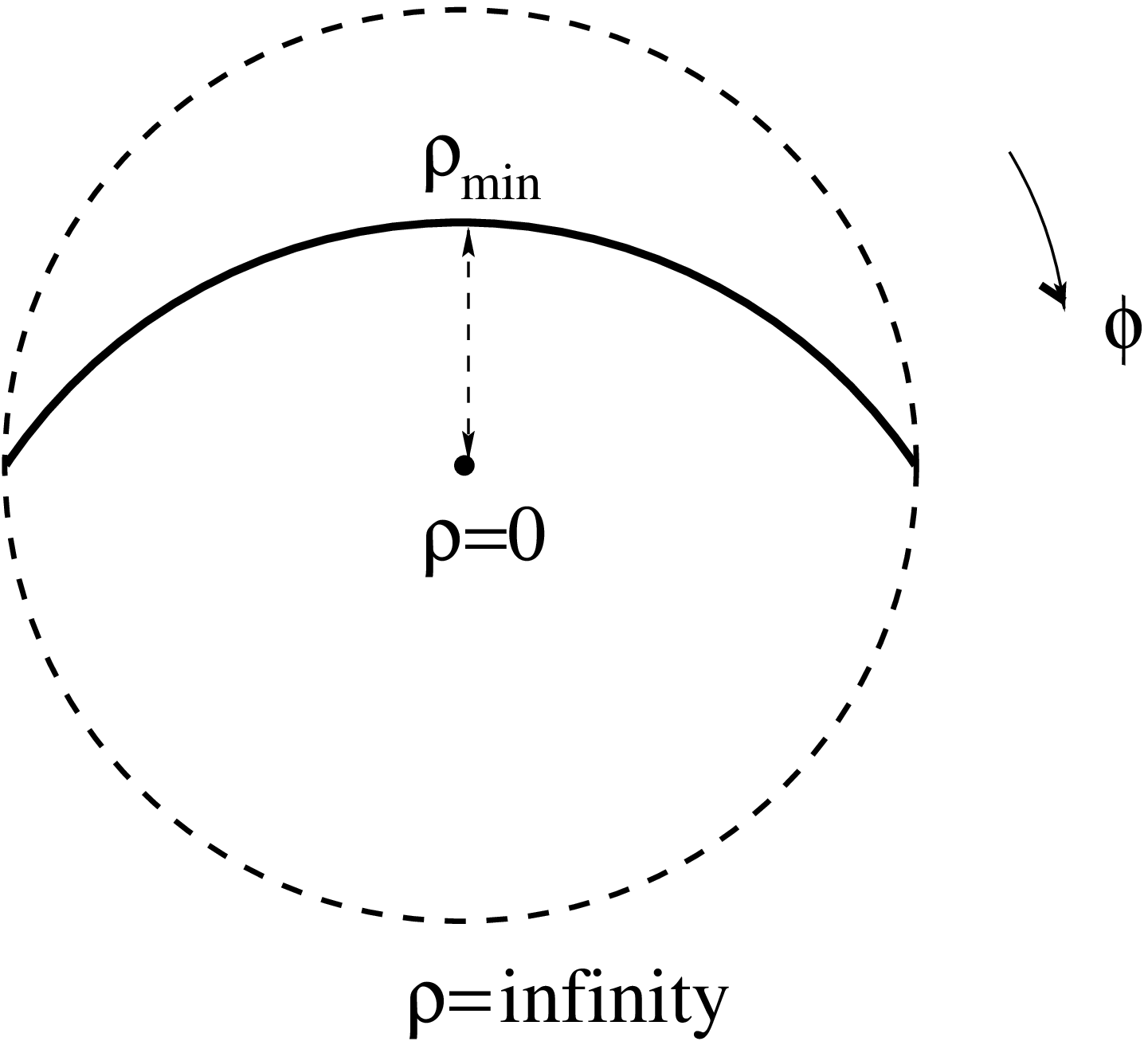,height=6cm}}
\end{picture}
\caption{A  $(1,q)$ string stretching between two antipodal
points on the boundary of
AdS$_3$ space. The minimal radius of the D-string is
$\rho_{\rm min} = \arcsinh ({qT_{\rm F}/ T_{\rm D}})$.
\label{vert}}}

Starting with the solution \eqref{adscon1} we can obtain
other classical  trajectories by acting
with non-trivial $SL(2,\bR)\times SL(2,\bR)$
transformations. These are generated   
(\romannumeral1) by $\phi$-rotations and  
(\romannumeral2) by boosts in the  
$(X^0, X^2)$ and $(X^2, X^3)$
planes of the embedding $\bR^{(2,2)}$ space.
World-volumes ``boosted" in the plane $(X^0, X^2)$ are defined   by the
relation
\begin{equation}
\cosh \beta\,\sinh \rho \, \sin \phi -
 \sinh \beta\, \cosh \rho\,  \cos \tau = {C} \, , 
\end{equation}
with  $\beta$ an arbitrary boost parameter.
They  describe periodic motions of the stretched string, with
a period $\Delta \t = 2 \pi$. The two end-points  in particular, at 
\begin{equation}
\phi = \arcsin \left( \tanh\beta \, \cos \tau\right) ,
\end{equation}
oscillate around the positions $\phi=0,\pi$.
All these trajectories are equivalent in AdS$_3$, since they
are related by exact isometries of the space-time.
We will thus focus our attention on the static situation  in what follows.

It will be useful  later to  visualize  the AdS$_2$
world-volume  in  Poincar{\'e} coordinates, in which the background
metric 
takes the well-known form: 
\begin{equation}
ds^2 = L^2 \left[{du^2\over u^2} +   u^2  \left(dx^2 -
dt^2\right)\right] .
\end{equation}
The Poincar{\'e} coordinates do not cover the entire AdS$_3$
(see Appendix A for details), but they
arise naturally in the near-horizon geometry of stringy black holes.
They are related to the  cylindrical coordinates as follows:
\begin{equation}
x \pm   t = {1\over u}\left(
\sinh \rho \, \sin \phi \pm \cosh \rho \, \sin \tau
\right)
\end{equation}
and
\begin{equation}
 u = 
\cosh \rho  \, \cos \tau + \sinh \rho  \, \cos \phi \,  .
\end{equation}
The  boundary of AdS$_3$ is at $\vert u \vert \to \infty$, while
$u=0$ is
an event horizon of the Rindler type. A static  observer in the center of
the cylinder appears
to be falling  through this event horizon, in
Poincar{\'e}  coordinates.

\vskip 3cm  
\FIGURE{\begin{picture}(300,150)(0,0)
\put(50,0){\epsfig{file=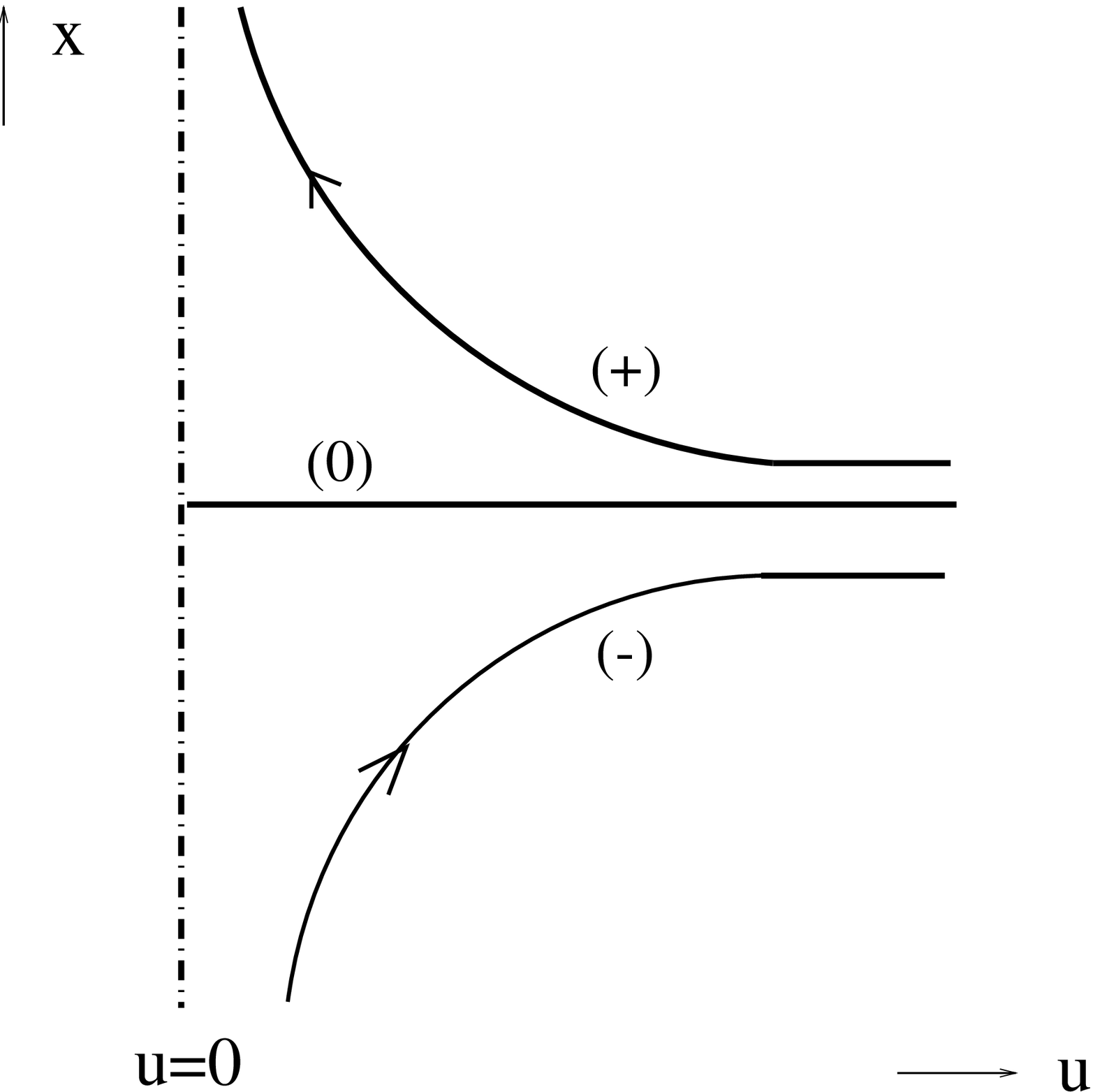,height=8cm}} 
\end{picture}
\caption{Static stretched $(1,q)$ strings,  with AdS$_2$ world-volume,
in the Poincar{\'e} coordinates.
The three cases correspond to $ C$ positive, zero and negative, as
explained
in the text. The arrows show the orientation of the bound fundamental
strings.
\label{vert1}}}

Our  static stretched string, Eq. \eqref{adscon1}, 
on the other hand, is also static
in the new coordinate system,  where it is given by
\begin{equation}
u   = { C\over x} \, .
\label{adscon2}
\end{equation}
This describes a string extending  radially in
towards the horizon at large $u$.
For $ C= 0$ the  string is straight and
cuts the horizon at  $x=0$, while 
for  $ C\not= 0$ it 
turns around and only approaches  the horizon asymptotically,     
see  Figure 3. In both cases the  length
of the string diverges near the horizon, but thanks to the
infinite redshift the tensive
energy  in this region stays finite.
It is worth noting  that  $\phi$-rotations of Eq. \eqref{adscon1}
give D-strings  that do not look static to the particular
Poincar{\'e} observer. 

Let us verify now that these static strings solve the
DBI  equations of motion. We will keep working
in Poincar{\'e} coordinates --
the analysis in global coordinates 
is presented  in  Appendix B. 
The general D-string configuration  in Poincar{\'e} coordinates
can be described by a 
function $u(x,t)$ and by a world-volume gauge-field strength $F_{xt}$. 
The induced metric is
\begin{equation}
\hat g = \left({L\over u}\right)^2 \left( 
\begin{array}{lll}
&{\dot u}^2 - u^4    & 
\;\;\;\;{\dot u}{u^\prime}  \cr &
\;\;\;{\dot u}{u^\prime}   
&{u^\prime}^2 + u^4  \cr
\end{array}
\right),
\end{equation}
where $\dot u\equiv \partial_t u$ and  $ u^\prime\equiv \partial_x u$.
Furthermore, the world-volume electric field
can be written as
\begin{equation}
{\cal F}  = L^2 \left(u^2 + f\right)\,  dx\wedge dt \, ,
\end{equation}
where $f\equiv 2\pi\alpha^\prime F_{xt}/L^2$, and we have
chosen the convenient gauge for the Neveu--Schwarz background:
$B = L^2 u^2\,  dx\wedge dt $ 
(recall that
$dB$ must equal  $2/L$ times the normalized
volume three-form). Putting all this together leads to
 the following DBI action
for a D-string:
\begin{equation}
I = - T_{\rm D} L^2 \int dt\, dx \, \sqrt{ u^4 + {u^\prime}^ 2 - {\dot
u}^2
- \left(u^2 + f\right)^2 }\, .
\label{BDI1}
\end{equation}
We are interested in  static extrema  of this action. 

The equation of motion for 
the gauge field  is the Gauss constraint,  which ensures the
continuity of electric flux: 
\begin{equation}
{2\pi\alpha^\prime \, T_{\rm D} \, {\cal F}_{xt}\over \sqrt{-
\det \hat g - {\cal F}_{xt}^2 } } = -q \in {\bf Z} \, .
\label{quant2}
\end{equation}
 Strictly-speaking, the Gauss constraint is the
$x$-derivative of Eq. \eqref{quant2}. We have written it in integral form  so
as to exhibit explicitly the fact that  $q$, which is
the number of fundamental
strings bound to the D-string,
 is quantized. Notice that the quantization argument given
in the previous section was based on the fact that the Wilson line
around   a closed string
is  periodic.  A piece of string does not ``know", however,
 whether it 
will eventually close or not. 
Locality, therefore, demands that the quantization be more generally
valid.
Notice  also that the Gauss condition
is equivalent to   Eq. \eqref{derived} -- a useful
rewriting for the manipulations that follow. 

To  solve  the remaining $u$-field equation we use
the continuity equation of the energy--momentum tensor $\Theta $;
because of the explicit two-dimensional Poincar{\'e}
invariance, this tensor is conserved. The (improved) energy--momentum
tensor is
\begin{equation}
\Theta ^\alpha_{\; \beta}
=  {\partial {\cal L} \over \partial
 \partial_{\alpha}u} \,  \partial_{\beta} u +
{\partial {\cal L}\over \partial F_{\alpha\gamma}}\, F_{\beta\gamma}
 - \delta^\alpha_{\; \beta}\, {\cal L}\, , 
\end{equation}
where the Greek indices run over $(t,x)$. 
For a static
string,   $\dot u=0$, 
the energy--momentum tensor is diagonal  so that, by the continuity
equation, 
$\Theta ^x_{\; x}$ 
is  a world-sheet constant.
 A somewhat lengthy but straightforward calculation, with the
help of Eqs. \eqref{derived} or \eqref{quant2} gives:
\begin{equation}
\Theta ^x_{\; x} =  L^2  \left(
  {T_{(1,q)}\, u^{4} 
\over \sqrt{u^4+ {u^\prime}^2 }} - qT_{\rm F}\, u^2  \right) .  
\end{equation}
Suppose  that $x$-momentum does not flow out of the string
 at infinity -- this amounts to  free boundary conditions in the
direction of the event horizon. 
We  must then demand that  $\Theta ^x_{\; x}=0$, which 
 is possible only for
$q\ge 0$,
and has  the general solution 
\begin{equation} 
u =  {C\over x-x_0} \ , \ \ {\rm with} \ \ 
{ C} = \pm {qT_{\rm F}\over T_{\rm D}}\, .  
\end{equation}
We have thus found the AdS$_2$ branes, 
 Eq. \eqref{adscon2}, up to an overall
translation  in  the $x$ direction. 
The constant $ C$, which determines
the radius of AdS$_2$, is  proportional to the number of fundamental
strings in the bound state. It is quantized in units of the
string coupling constant $\lambda_s$ -- this  is invisible in the CFT
limit.  The AdS$_2$ radius is equal to 
\begin{equation} 
\ell_{{\rm AdS}_2}  = L\sqrt{1+{ C}^2}= L\, { T_{(1,q)}\over 
T_{\rm D}} \, .
\label{radius}
\end{equation}
The fact that $q$ is positive
 implies a definite orientation for the fundamental
strings, as shown in Figure 3. Notice finally that
$F_{xt}=0$
for these solutions --  gauge-invariant meaning
can, however, only be attached to ${\cal F}_{xt}$.

Allowing an arbitrary $\Theta ^x_{\; x}$ leads to a more general
class of static solutions. 
Some of these  describe strings that
return to the boundary of
AdS$_3$  without crossing the  event horizon at $u=0$. 
In the holographically-dual CFT  string end-points probably  
correspond to heavy external sources, analogous to the external quarks
in $N=4$ SYM. 
The above solutions  would
enter  the calculation of the static force between such 
 external sources.
 Supersymmetry could  constrain
the allowed boundary conditions, probably along the lines of Ref. 
\cite{DGO}. We will not  pursue these interesting issues
 here any further.

The  energy of our static string,
as measured by an observer sitting at radial position
$u=1$ in the AdS$_3$ space, is
\begin{equation}
E  = {1\over L}\int dx\, \Theta^t_{\; t} =   \int dx\,  L 
\left( T_{(1,q)} \sqrt{u^4+ {u^\prime}^2} - qT_{\rm F}\,  u^2
 \right)\, .
\end{equation}
This is  the sum of  tensive energy plus 
the interaction with the $B$-field background.  
The reader is invited to check that $u={ C}/x$ solves indeed the
local minimization condition, as advertised.
 The total energy 
diverges  near the boundary of AdS$_3$, but it
 is on the other hand convergent
 near the event horizon. A straightforward calculation with the help
of Eq. \eqref{radius} gives:
\begin{equation}
E = T_{\rm D} L u_0 \, ,
\end{equation}
where $u_0$ is a cut-off in the radial coordinate.
Note that the number,  $q$, of fundamental strings
has dropped out of the above expression.

Our final  comment concerns the extension
of this solution behind the event horizon, at  $u=0$. 
As seen in Figure 2, and discussed further in Appendix B,  
an  observer at the center of the cylinder sees a static string
stretching across antipodal points on the boundary -- nothing
special happens,  from his point of view,  at the Poincar{\'e}  horizon. 
To a Poincar{\'e} observer, on the other hand, the string seems to extend
along the horizon indefinitely (for $q\not= 0$) without ever crossing
inside. This  is analogous to the well-known phenomenon
that a particle trapped by a black hole
appears to be  falling in 
eternally in  the eyes of an  outside observer. 


\boldmath
\section{Supersymmetric AdS$_2\times S^2$  branes}
\unboldmath

The stability of the AdS$_2$ branes would be garanteed 
if we can embed them 
in a setting in which some supersymmetries are preserved.
One such setting is provided by
the near-horizon geometry of a black string,
constructed out of $Q_5$
NS five-branes and $Q_1$ fundamental strings  of  type-IIB  theory.
Both of these extend along the  non-compact direction  $x$, while the 
five-branes wrap also a compact four-manifold ${\cal M}_4$
(e.g. a four-torus,  or K3). 
The near-horizon geometry of this configuration
is  AdS$_3 \times S^3 \times {\cal M}_4$. The AdS$_3$ is parametrized
by the
coordinate $x$, the  time coordinate $t$, and the
radial coordinate measured from the horizon $u$.
We will take for simplicity ${\cal M}_4 = T^4$ in what follows.
This near-horizon geometry   admits an exact CFT description, as a
supersymmetric WZW model on the product manifold
  $SL(2,\bR)\times SU(2)\times 
U(1)^4$.
The radii  of the AdS$_3$ 
space  and of $S^3$  are equal to each other,  and fixed by
the number  of NS five-branes,
\begin{equation}
L^2 = Q_5\, \alpha^\prime = (k+2) \alpha^\prime\, .
\label{exact}
\end{equation} 
Here $k$ is the level of the $SU(2)$ bosonic current algebra, while the
level of the $SL(2,\bR)$ algebra is $\tilde k = -(k+4)$. One can check
that the total central charge, taking also into account the 
$U(1)^4$ torus part, adds up to the critical value $c=3/2\times 10$.
The above expression for the radii, valid 
a priori in the weak-curvature limit, is protected by  the unbroken
supersymmetry and is expected to be exact.

A simple set of D-branes in this geometry can be obtained by putting
together the stretched D-strings  of the previous section, 
and the spherical  D2-branes of the
WZW model for  $SU(2)$.
 Let us recall briefly some
salient features of these  latter \cite{as,bds}.
 The  regular conjugacy classes
of $SU(2)$  are two-spheres embedded in the
 three-sphere group manifold. Since $\pi_2(S^3)$ is trivial, the corresponding
 branes
could  a priori shrink to zero size. What stabilizes them at a fixed radius
is a world-volume magnetic flux, quantized in units
 of the inverse area of the brane. There is one D-brane for each integer
$0< p \alpha^\prime < L^2$, and it 
carries $p$ units of magnetic flux. The induced metric and 
the gauge-invariant
two-form (${\cal F}= \hat B + 2\pi\alpha^\prime F$) for the $p$th
brane read:
\begin{equation}
ds^2_{\vphantom 2} = L^2_{\vphantom 2}\,  
\sin^2_{\vphantom 2}\left({\pi p \alpha^\prime\over L^2}\right)
d\Omega_2^2
\ \  {\rm and} \ \  
{\cal F} = - {L^2\over 2} \, \sin\left({2\pi p \alpha^\prime \over
L^2}\right)
d\omega_2\, ,
\label{radius1} 
\end{equation}
where $d\Omega_2^2$ and ${d\omega}_2^{\vphantom 2}$ are the 
conventionally-normalized
metric and the corresponding  volume
form on the two-sphere (not to be confused, hopefully,
 with the group
automorphism of  the gluing conditions!).
The  D-brane mass and charge,
derived from  the Dirac--Born--Infeld and the Wess--Zumino actions, as
well
as the spectrum of small fluctuations, 
agree as should be expected with the CFT results in the semi-classical, 
large-$k$ limit. More surprisingly,  the agreement is in fact exact,
if one takes Eq. \eqref{exact} at face value for all $k$ \cite{bds}.

Combining now these spherical D2-branes with the D-strings of
Section 4, 
gives a set of  D3-branes in AdS$_3 \times S^3$,  with
induced geometry  AdS$_2 \times S^2$.
Such branes  correspond to  conformal boundary states, since they
can be  obtained by tensoring (in
the Neveu--Schwarz  and Ramond  sectors separately) conformal 
boundary states for
$SL(2,\bR)$ and $SU(2)$. Alternatively,  one can use the factorization
of the DBI action  to prove that the semi-classical  equations
admit solutions of such product form.  The only subtlety concerns the
quantization condition \eqref{quant2}, whose left-hand-side
must be now multiplied by an extra factor of $p$. 
Accordingly, the constant $C$ and the AdS$_2$ radius are
modified to 
\begin{equation}  
{ C} = \pm {qT_{\rm F}\over p T_{\rm D}}\, \ \
{\rm and }  \ \ \  
\ell_{{\rm AdS}_2} = L\, { T_{(p,q)}\over p T_{\rm D}} \, .  
\end{equation}

An  AdS$_2 \times S^2$ brane
is drawn schematically
in Figure \ref{d3}: it is a  (hyper)tube of  fixed spherical cross section,
which  approaches  tangentially the horizon of the background 
NS5/F1 black string. The brane carries the quantum numbers of a $(p,q)$
string, as can be checked by calculating the world-volume fluxes through
the two-sphere. 
Notice that the  quantized  fluxes  are the integrals
 (\romannumeral1) of  the
magnetic field  $F = dA = -\frac{p}{2}\,  d\omega_2$, rather than of the
gauge-invariant combination  ${\cal F}$,   and
(\romannumeral2) of
 the (dual)  electric displacement 
$\star{\partial {\cal L}/ \partial F_{xt}}$.  
This is consistent with the fact that
the elementary electric charge -- the
end-point  of a fundamental string --   couples minimally  to the
gauge potential $A$.

\vskip 1.5cm
\FIGURE{\begin{picture}(300,150)(0,0)
\put(10,0){\epsfig{file=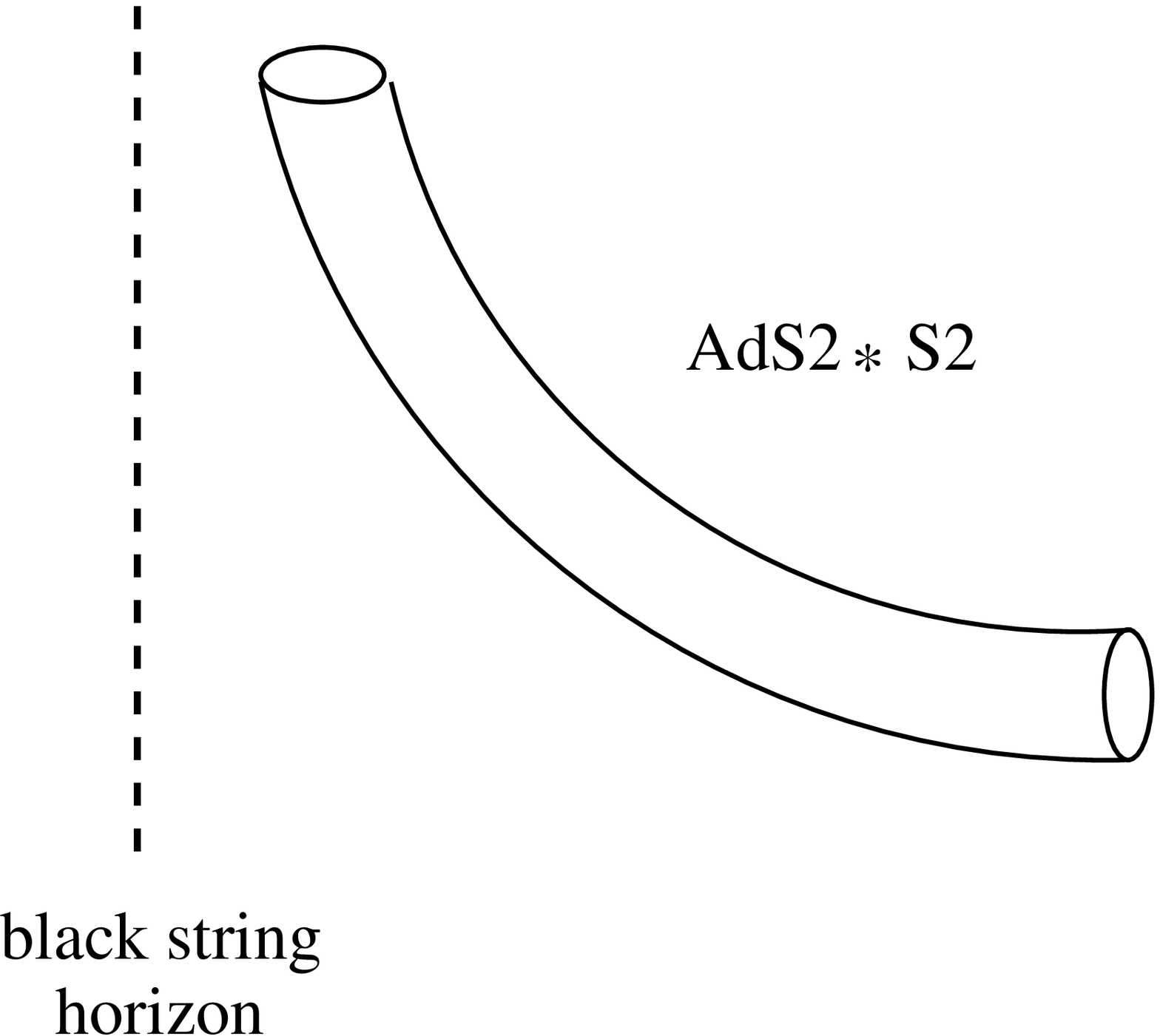,height=7cm}}
\end{picture}
\caption{The  AdS$_2 \times S^2$ D-brane in the near-horizon geometry of
a NS5/F1 black string. The D3-brane carries the quantum numbers of a
$(p,q)$ string.
\label{d3}}}

The NS5/F1  black string background preserves 1/4 of the 32 type-II
supersymmetries. These are doubled in the near-horizon geometry,
where the background invariance is enhanced to superconformal. 
We want to show now that half of the unbroken  near-horizon
supersymmetries  are also preserved by the   AdS$_2 \times S^2$ D-branes.
This is easiest to establish 
 directly in  the CFT, by extending  the arguments
of Refs. \cite{abs} and \cite{sta1}. 
The  supersymmetric WZW model  has  ten  
free fermionic world-sheet coordinates, transforming in 
the adjoint representation of the product group. 
Let us call the left- and right-moving fermions 
$\psi^A$ and $\bar\psi^A$, with $A$
a flat tangent-space index.  The unbroken space-time supersymmetries 
must obey   the usual GSO projections:
\begin{equation}
\left(\prod_{{\rm all \ } A} \Gamma^A \right) Q = Q 
\ \ {\rm and} \ \ 
\left(\prod_{{\rm all \ } A} \Gamma^A \right) \bar Q = \pm \bar Q \, ,
\label{chi}
\end{equation}
where $Q$ and $\bar Q$ are two ten-dimensional Weyl--Majorana
spinors, and the minus or plus sign refers to  type IIA or type IIB.
In addition, the non-trivial background imposes two extra chiral projections:
\begin{equation}
\left( \prod_{A\notin U(1)^4} \Gamma^A \right)Q = Q  
\ \ {\rm and} \ \ 
\left (\prod_{A\notin U(1)^4} \Gamma^A\right) \bar Q =  \bar Q \, ,
\label{chir}
\end{equation}
where only the $SL(2,\bR)\times SU(2)$ tangent indices enter in the product. 
From the world-sheet point of view, these
projections follow from the super-Virasoro conditions, because of 
the trilinear terms in the superconformal generators.
These  projections  reduce the space-time supersymmetry by a half, 
as expected.

Now our  AdS$_2\times S^2$ branes impose  the
gluing conditions on  world-sheet boundaries:
\begin{equation}
J^A = -\omega \left(\bar J^A \right) 
\ \ {\rm and} \ \ 
\psi^A = -\omega \left(\bar \psi^A \right)\, ,
\end{equation}
with $\omega$ the corresponding  algebra automorphism. 
Consistency of the operator product expansions implies  appropriate
boundary conditions on spin fields, such  that any  unbroken
supersymmetries must be of the form \cite{revs}
\begin{equation} 
Q + \Omega  \bar Q \, ,
\label{unbr}
\end{equation}
with  $\Omega$  the action of the automorphism  
on the $SO(1,9)$ spinors. To be more explicit,  $\omega$ induces  an
automorphism of the $SO(1,9)$ algebra in spinor space, which can
be implemented by conjugation with an element of the group:
\begin{equation}
\Sigma^{AB} = {i\over 4} 
\left[\Gamma^A,\Gamma^B\right] \to
{i\over 4} \left[\omega \left(\Gamma^A\right),
\omega\left(\Gamma^B\right)\right] 
\equiv \Omega\, \Sigma^{AB}\, \Omega^{-1} \, . 
\end{equation}
This defines the spinor-transformation matrix $\Omega$.
Now since
 $\omega$ does not mix the $SL(2,\bR)\times SU(2)$ and $U(1)^4$ currents,
it follows  that $\Omega$ acts as a tensor product
 on the corresponding $SO(1,5)$ and $SO(4)$ spinors. Thus, $\Omega$ 
is compatible with  the
 chirality
projections \eqref{chir}, and hence the unbroken 
supercharges \eqref{unbr} can be defined. This proves the supersymmetry
of the AdS$_2 \times S^2$ branes, as advertised.

\vskip 1.5cm
\FIGURE{\begin{picture}(300,150)(0,0)
\put(10,0){\epsfig{file=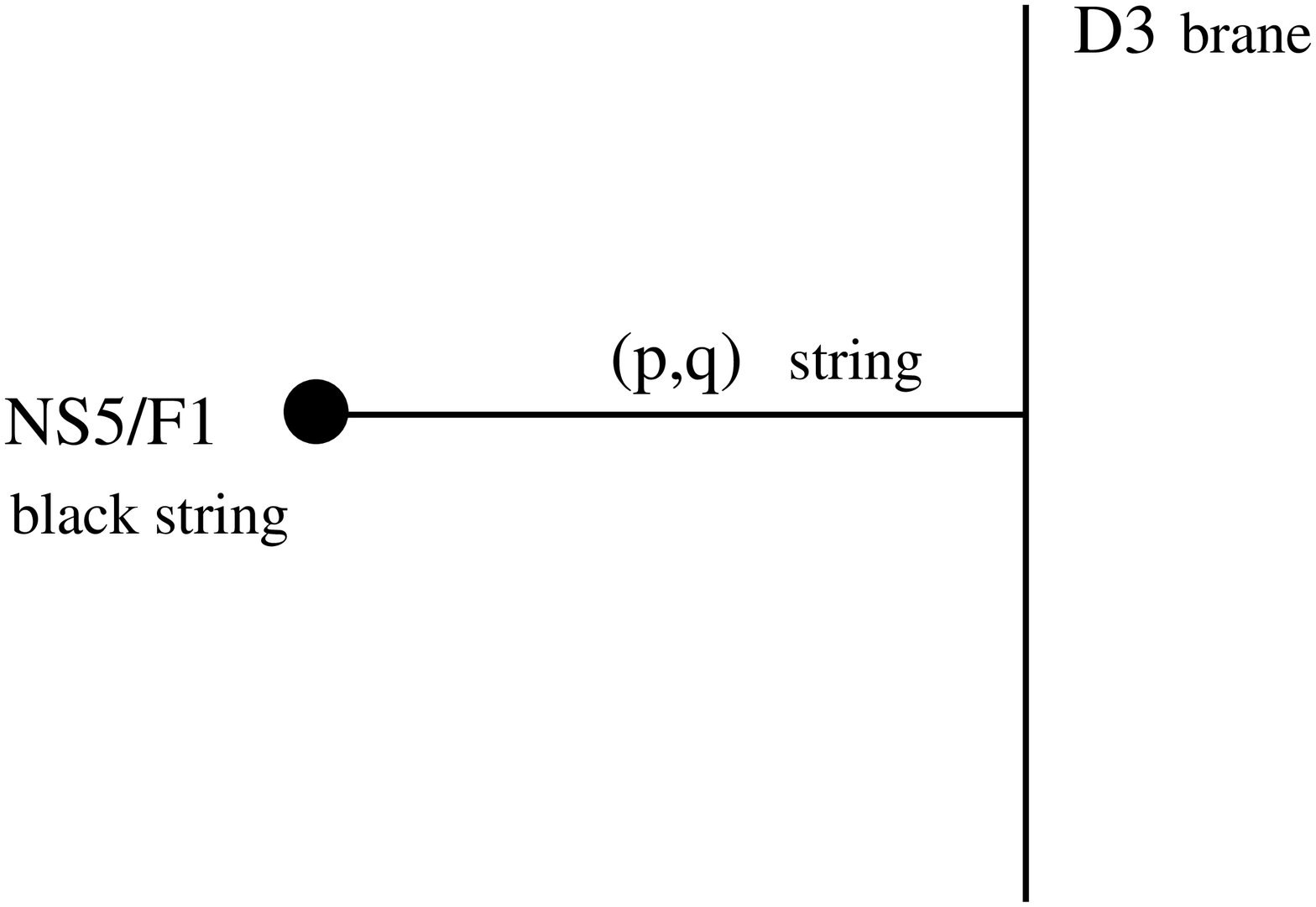,height=6cm}}
\end{picture}
\caption{A  $(p,q)$ string stretching between a flat D3-brane and
an orthogonal  NS5/F1 black string. The AdS$_2$$\times $$S^2$ branes
describe the $(p,q)$ string
 in the near-horizon region of the black string.
\label{rn}}}

Whether some of these supersymmetries can be preserved in the
asymptotically-flat region of the black string is less clear. 
Figure \ref{rn} illustrates a 
$(p,q)$ string emerging  from   the NS5/F1 black string,  and 
ending, for example, on an orthogonal distant  D3-brane  (this latter
 plays no role in  our argument  here). 
The $(p,q)$ string and the orthogonal  F-strings and NS5-branes
have mutually-incompatible supersymmetries, so one may
hastily conclude that the configuration is not supersymmetric. 
This need not,  however, be a priori true. If $q=0$ for example,
the D-strings and the orthogonal NS5-branes break only 1/4 of  the
supersymmetries. Furthermore, the $p$ D-strings can merge with
the $Q_1$ background  F-strings  into a $(p, Q_1)$ bound state  at a supersymmetric
junction \cite{jun1,jun3,jun2}. 
If   $Q_1$ is much  greater
 than  $pT_{\rm D}/T_{\rm F}$, the angle
at the junction is almost 
equal to $\pi$, so that the  F-strings will  be almost  straight. 
Thus it is very likely that some of the near-horizon supersymmetries
survive in the asymptotic region in this case. 
It would be interesting to study this question in more detail.


\section{The view from the brane}

The above CFT proof of supersymmetry 
appears at first sight  to contradict the following, alternative  
argument: 
the effective theory of the $U(1)$ vector multiplet on the 
D3-brane has
a (global)  supersymmetry, if and only if
the background  geometry 
admits  a covariantly-constant spinor\footnote{To see why  
start with  the $N=1$ supergravity plus super-Maxwell
Lagrangian in four dimensions,
and freeze the  supergravity fields to some background values. Rigid
supersymmetries of the resulting theory should not
transform the supergravity backgrounds. But for
the gravitino field to stay inert,
the supersymmetry parameter must be a covariantly-constant spinor.}.
This in turn implies that the Ricci scalar curvature must vanish, which
in our case means that the AdS$_2$ and   $S^2$ radii 
$\ell_{{\rm AdS}_2}$ and $\ell_{S^2}$,  respectively) must be  equal.
From our previous discussion, however, it follows that
\begin{equation}
\ell_{{\rm AdS}_2}= L \, {T_{(p,q)}\over pT_{\rm D}} \geq   L  \geq 
\ell_{S^2} =
L \sin\left({\pi p \alpha^\prime\over L^2}\right) . 
\label{radii} 
\end{equation}
The  two radii are thus  only  equal  in the special case  $q=0$ and
$p =L^2/2\alpha^\prime$, whereas  our CFT proof of  supersymmetry was valid
for all  values of $p$ and $q$. Clearly  one of the two reasonnings
must be  wrong.

The fallacy is actually in  the above effective-field-theory argument. 
The radii
\eqref{radii} were calculated with  the {\it induced} world-volume metric,
but
the vector-multiplet states are  open strings  which
couple  to an {\it effective open-string} metric.  It is  this latter
metric 
that must  admit a covariantly-constant spinor. 
Since the ${\cal F}$-field  on the D3-brane is a closed two-form,  
the induced and open-string  metrics are related by
the standard formula \cite{open1}:
\begin{equation}
G_{\alpha\beta} =
\hat g_{\alpha\beta} - 
{\cal F}_{\alpha\gamma}\, 
\hat g^{\gamma\delta}\, 
{\cal F}_{\delta\beta}\, .
\end{equation}
Both ${\cal F}$ and the metric  are block-diagonal,
 so we may  compute the 
right-hand side separately for the AdS$_2$ and $S^2$
components of the brane.  A simple calculation, 
using Eqs. \eqref{radius1},  gives for the two-sphere:
\begin{equation} 
ds^2_{\rm open} =  L^2_{\vphantom 2} \,d\Omega_2^2\, .
\label{twospher}
\end{equation}
Similarly, from the results of the previous section, one can calculate readily
the induced metric and invariant field on AdS$_2$:
\begin{equation}
ds^2 = L^2
\left(1+C^2\right)
{du^2\over u^2} - L^2  u^2  \,dt^2 
\ \ {\rm and} \ \ 
{\cal F} = L^2~  C \, dt \wedge du \, , 
\label{criti}
\end{equation}
from which one derives easily:
\begin{equation}
ds^2_{\rm open} =  L^2 \left[ {du^2\over u^2} -u^2 \, d{\tilde t}^2 \right] 
\ \ {\rm with}\ \ {\tilde t} = {t\over \sqrt{1+C^2}}.
\end{equation}
Thus, to  an observer on the D3-brane, the radii of  $S^2$ and AdS$_2$ 
appear  equal to each other, and to the  radius  $L$ of the ambient geometry, 
independently of the  quantum numbers  $p$ and $q$.
This is precisely what the   unbroken  supersymmetry requires.

 We find the above result
 quite remarkable,  and suspect that it could
 be more generally valid. Notice that to  an observer in the
bulk the  two-sphere may look exceedingly  small,
\begin{equation}
\ell_{S^2} \ll 
L\ \  {\rm  if}\ \  p\alpha^\prime/L^2\ll 1
\label{limit1}
\end{equation}
 and the
AdS$_2$ space arbitrarily flat,
\begin{equation}
\ell_{{\rm AdS}_2} \gg L\ \ {\rm  if}\ \ 
qT_{\rm F}\gg pT_{\rm D}.
\label{limit2}
\end{equation}
Yet, from the perspective of a  brane observer
 the radii cannot vary at all -- they are frozen by  the ambient
curvature! 
This could be also deduced  from the  CFT:  the 
conformal weights of open-string
vertex operators
are the same as those of closed-string vertex operators, and
do not depend on the quantum numbers of the D-brane, 
 $p$ and $q$. These latter  enter in the choice of allowed
representations, but do  not therefore  affect the
covariant wave operator on the brane.
 The fact that  
the AdS$_2$ brane stays effectively curved, even though its
induced metric is almost flat, 
could be significant 
for Randall--Sundrum  compactifications of string theory\footnote{
The spherical $SU(2)$ branes may be of interest also in  more
conventional compactifications, similar in spirit to those of
 references
\cite{b,bgkl}.}.
Notice, in passing,   that the open-string 
geometry is identical to that of an extremal dyonic (Reissner--Nordstr{\"o}m)  
four-dimensional black hole. Note also  that 
the propagation of open strings does  not violate 
causality in the bulk, as is  expected on  general grounds \cite{gary1}.

 The world-volume theory on the D3-brane
is particularly interesting in the limits
\eqref{limit1} and \eqref{limit2}. Consider the limit \eqref{limit1}
first: from Eqs. \eqref{radius1}  it follows easily that
${\cal F}\gg \hat g$, so that the open strings will  behave
as light  dipoles 
in a  strong external   magnetic field.
Each string endpoint, in particular,  is localized
on one of the $p$ lowest Landau states, so that the string has $\sim p^2$
low-lying excitations. 
This is precisely
the decoupling limit in which  the effective
low-energy theory is a non-commutative gauge theory on the fuzzy
two-sphere.  
Consider next the limit \eqref{limit2}, which implies $C\to\infty$.
From Eqs. \eqref{criti}  we conclude  that the electric field tends
to its critical value in this case, so the world-volume theory is
a curved analog of the conjectured non-commutative theory of
open strings \cite{sst,gmms}. The presence of any nearby D-brane may
lead in this case to infinitely-fast dissipation \cite{bp,dyn}.
We leave these questions to  future investigation.


\section{D-branes in the BTZ geometry}

In this final section we  discuss  the embedding of 
our  D-branes in the background of a  BTZ black hole.
This latter 
is part of  the near-horizon geometry for the general
static five-dimensional 
black hole, which has  $Q_5$ NS5-branes, $Q_1$ fundamental strings and 
$N=N_{\rm L}-N_{\rm R}$ units of Kaluza--Klein momentum 
in the (compactified) string  direction. 
Since the BTZ geometry can be  obtained by periodic identification of (a 
part  of)  the universally-covered AdS$_3$,  the  D-branes discussed
previously are  still
solutions of  the  DBI equations. Hence, we need only worry about
(\romannumeral1) their global structure,  and 
(\romannumeral2) their physical interpretation  in
this  novel setting.  Euclidean signature or a  supercritical 
world-volume field are ``pathologies"  at the local level,   
so the $H_2$ and dS$_2$ solutions
remain ``unphysical".  We will thus  concentrate on  D-branes which are
locally AdS$_2$. 

Let us first recall 
certain standard facts about the BTZ geometry \cite{btz}.
This  is obtained after  modding out  AdS$_3$  by a  discrete 
isometry
\begin{equation}
\exp 2\pi \xi \equiv \exp 2\pi \left( r_+ \, \xi_{32} -  r_- \,
\xi_{01}\right)
\, , 
\label{discrete}
\end{equation}
where  $\xi_{MN}$ are generators of Lorentz boosts
in the $(X^M,X^N)$ plane
of the embedding Cartesian space. Without loss of generality we
may assume
$r_+\geq r_-\geq 0$.  To avoid closed time-like curves,
we must insist that the norm of the Killing vector $\xi$ 
(defined in an obvious way) stays everywhere
positive. Negative-norm regions of AdS$_3$ must thus be removed 
before  identifications.   For instance, if $r_- =0$ 
one must keep only the interior of the half light cone 
in  the $(X^3,X^2)$ hyperplane.
The boundaries of the excised  regions are
singularities in the causal structure
of the BTZ black hole. In the regions where  $\xi$ is space-like,
we  may choose coordinates such
that
\begin{equation}
\partial_{\varphi} \equiv \xi \ , \ \ 
\partial_{t^\prime} \equiv  
r_+\, \xi_{01} - r_-\, \xi_{32} \ \ 
{\rm and} \ \  
r^2 \equiv (\xi, \xi) \, .
\end{equation}
These  are  related to the  coordinates  $X^M $ as follows: 
\begin{equation}
X^0 \pm X^1 =
\left\{
\begin{array}{ll}
\pm \epsilon_1
L \left(r^2_{\vphantom+}- r^2_+
\over r^2_+ - r^2_-\right)^{1/2}  \exp \pm 
\left(r_+ {t^\prime} - r_- \varphi \right)
\ ,& \ \ {r>r_+}\; ,\cr
{\hphantom +} \epsilon_2
L \left(r^2_+ - r^2_{\vphantom+}
\over r^2_+ - r^2_-\right)^{1/2}  \exp \pm 
\left(r_+ {t^\prime} - r_- \varphi \right)
\ ,& \ \ {r<r_+}\cr
\end{array}   
\right.
\label{tr0}
\end{equation}
and
\begin{equation}
X^2 \pm X^3 =
\left\{
\begin{array}{ll}
\pm \epsilon_3
L \left(r^2_{\vphantom+}- r^2_-
\over r^2_+ - r^2_-\right)^{1/2} \exp \pm 
\left(r_+ \varphi - r_- {t^\prime} \right)
\ ,& \ \ {r>r_-}\; ,\cr
{\hphantom +}\epsilon_4
L \left(r^2_- - r^2_{\vphantom+}
\over r^2_+ - r^2_-\right)^{1/2} \exp \pm 
\left(r_+ \varphi - r_- {t^\prime} \right)
\ ,& \ \ {r<r_-}\; .\cr
\end{array}   
\right.
\label{tr1}
\end{equation}
Here the $\epsilon$'s  are pure signs -- several different choices
of them are necessary in order to  cover entirely  the
part  of a  hyperboloid where  $\xi$ is space-like.
Note also that the above transformations degenerate in the extremal
case  $r_+=r_-$. 

The metric in the  $(t^\prime,r,\varphi$)   coordinates has the
canonical  BTZ form:
\begin{equation}
ds^2 =L^2 \left[
-f^2(r) \, {dt^\prime}^2 + f^{-2}(r) \, dr^2 + r^2 \left(
d\varphi-{r_+ \, r_- \over   r^2}\, dt^\prime\right)^2\right]\,  ,
\label{btz1}   
\end{equation}
with
\begin{equation}
f(r) ={1\over r}\sqrt{\left( r^2_{\vphantom +}-r^2_{+} \right)
\left(r^2_{\vphantom -}-r^2_{-}\right)} \, . 
\label{btz2}   
\end{equation}
The discrete identification makes $\varphi$ an angular variable,
$\varphi \cong \varphi+2\pi$. The  BTZ geometry   describes a 
three-dimensional black hole,  with mass $M$ and angular momentum $J$, in 
a space-time that is asymptotically anti-de Sitter. The singularity at
$r=0$ is
hidden behind an inner horizon at $r=r_-$, and an outer horizon at
$r=r_+$.
Between these  two horizons,  $r$ is time-like.
The coordinate $t^\prime$ becomes space-like inside the ergosphere,
when $r^2_{\vphantom g}< r^2_{\rm erg} \equiv r_+^2 + r_-^2$. 
The relation between $M,J$ and  $r_\pm $ is as follows:
\begin{equation}
r_\pm ^2  = 
{ML \over 2}\left[
1\pm \sqrt{1-\left({J\over  ML }\right)^2}
\right] .
\end{equation}
Extremal black holes have  $\vert J \vert =ML$. In the special case $J= ML=0$
one finds the near-horizon geometry of the
five-dimensional  NS5/F1 stringy black hole in its ground state. Since
$r_+=r_-=0$, 
 the  BTZ and Poincar{\'e} coordinates ($u>0$) coincide in this special
case.
 The ground state of
the stringy black hole should be distinguished from global  AdS$_3$, 
because  the  region of negative  $r^2$   (where $\varphi$ 
would have been   time-like) is  excised, and because $r=0$ is a real
singularity. 
Global  AdS$_3$  is obtained for $J=0$ and $ML=-1$\footnote{This  arises 
in  the near-horizon
geometry of the (uncompactified) six-dimensional black string, but
also of certain special, spinning, five-dimensional black holes (see
\cite{mm} for a recent discussion).}. One can  check that 
$(t^\prime,r,\varphi)$ can be identified with the 
  global cylindrical 
coordinates  $(\t,\sinh\rho,\phi)$  
in this case.

Let us return  now to the AdS$_2$  D-branes
which are intersections of  the (multiply-covered) hyperboloid
with a hyperplane  $X^M=L C$, where $X^M$ is space-like.
Different choices for the  embedding coordinate $X^M$  
are  equivalent in global AdS$_3$, but not in  the BTZ geometry
where the discrete 
identifications break  the  $SL(2,\bR)_{\rm L}\times SL(2,\bR)_{\rm R}$
symmetry. Depending on the choice of $X^M$  
the D-brane 
may, in particular, either  avoid or hit the BTZ singularity. A  piece
of its  world-volume must  be, of course, excised in the latter case. 

\vskip 1.5cm
\FIGURE{\begin{picture}(300,150)(0,0)
\put(10,0){\epsfig{file=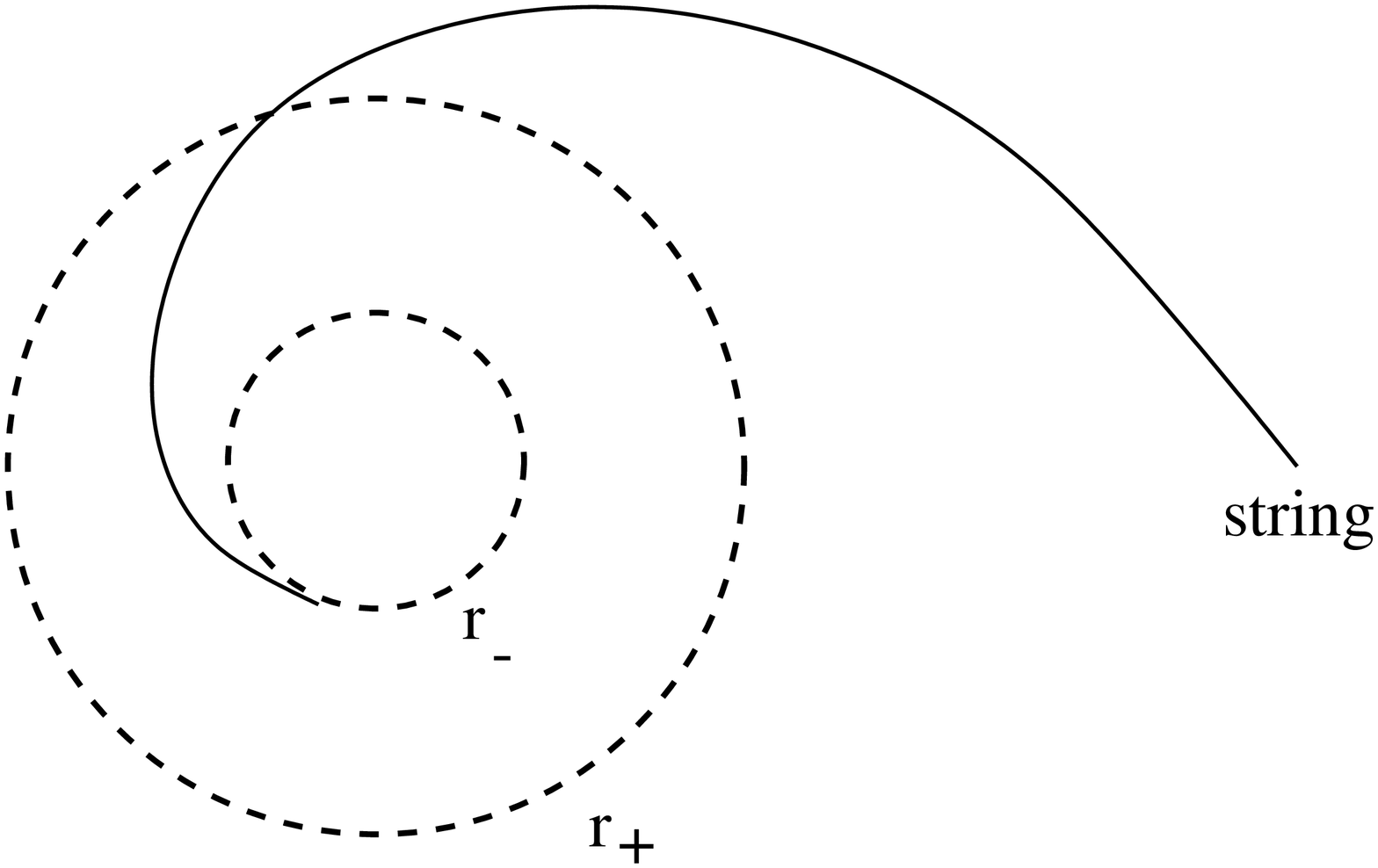,height=6cm}}
\end{picture}
\caption{A D-string in the BTZ geometry spiralling  in
past the outer horizon, 
and then circling  around the  inner horizon  an infinite
number of times. The completion of this  world-sheet
takes  it behind  the inner horizon, where it either hits the
BTZ singularity, or it reemerges 
 eventually on the outside. 
\label{xxx}}}
\vskip 0.3cm

Consider for example $X^2=L C$. Using
Eqs. \eqref{tr1} one finds,  in the region $r>r_-$:
\begin{equation}
\sqrt{r^2_{\vphantom -}-r_-^2}=
 {\epsilon_3 \, C \sqrt{r_+^2-r_-^2} \over \sinh\left (r_+\varphi
-r_-t^\prime\right)}\, . 
\label{prec}
\end{equation}
At fixed $t^\prime$, this describes
a $(1,q)$ string coming in from infinity, crossing the outer
horizon,  and then spiralling  around the inner
horizon an infinite number of times, see Figure 6. 
The world-sheet \eqref{prec} is  not
complete -- a  world-sheet cannot terminate elsewhere  other than at
the singularity,  or at the boundary of the space-time.
Continuing  behind the inner horizon, at  $r< r_-$, we find: 
\begin{equation}
\sqrt{r_-^2-r^2_{\vphantom -}}=
{\vert C\vert \sqrt{r_+^2-r_-^2}\over \cosh\left(r_+\varphi
-r_-t^\prime\right)}\, . 
\end{equation}
For $\vert C \vert\ge  r_- \big/ \sqrt{r_+^2-r_-^2}$, 
the world-sheet  hits the singularity,  at $r=0$, and terminates.
For smaller $\vert C\vert$, on the other hand,  
the string  avoids  the singularity  
and reemerges  on the outside. 
From there  it can be
continued back  the AdS$_3$ boundary. Since $C=qT_{\rm F}/T_{\rm D}$, 
it is  the number of fundamental strings in the $(1,q)$ bound state that 
determines  its fate. Notice that as
time goes on,  the string undergoes a 
$\varphi$-precession with constant angular velocity $r_-/r_+$
(for  $r_-=0$ it  is static).

For a different  type of D-brane consider the inequivalent choice 
$X^1= LC$. In the region  $r >r_+$ this constraint reads:
\begin{equation}
\sqrt{r_{\vphantom +}^2-r_+^2}=
{\vert C\vert \sqrt{r_+^2-r_-^2} \over \cosh\left(r_+t^\prime -r_ -
\varphi\right)}\, . 
\label{prec1}
\end{equation}
It describes again a spiralling and precessing string, which is
now contained  within a maximum radius, 
\begin{equation}
r_{\rm max} = \sqrt{r_+^2 + C^2_{\vphantom -} \left(r_+^2-r_-^2\right)} 
\, .
\end{equation}
In the special case $r_-=0$ the string is actually
finite and circular. It  shoots out from the outer horizon,  before
turning around  to  fall back  inside. 
Continuing its trajectory behind the  outer horizon we find:
\begin{equation}
\sqrt{r_+^2-r^2_{\vphantom +}}= {\epsilon_2 \, C \sqrt{r_+^2-r_-^2}\over
\sinh(r_+t^\prime -r_-\varphi)}\, . 
\label{prec2}
\end{equation}
Since the hyperbolic sine takes all posible values in $\bR$, there
is no way to  avoid the singularity in this case: the  $(1,q)$ string  
is necessarily doomed to hit it. 
 
In the BTZ geometry, our D-branes cannot be supersymmetric.
To see  why, express  the identification under the
discrete isometry \eqref{discrete} in terms of $SL(2,{\bf R})$ group
elements:
\begin{equation}
g \cong g_{\rm L}^{\vphantom 1} \, g \,g_{\rm R}^{-1} \, ,
\end{equation}
where  
\begin{equation}
g_{\rm L} = \exp{\pi\left(r_+-r_-\right)\sigma^3}
\ ,  \ \ 
g_{\rm R} = \exp{\pi\left(r_++r_-\right)\sigma^3}
\, , 
\label{hyper}
\end{equation}
and $\sigma^3$ is  the diagonal Pauli matrix. The 
 $g_{\rm L}$ ($g_{\rm R}$) act
as $SO(1,2)$ boosts  on the left-moving (right-moving) currents
of the WZW model, and on their fermionic partners. 
As a result, they also act as Lorentz boosts  on the spinor supercharges
$Q$ ($\bar Q$). Since there is no invariant spinor under a Lorentz boost,
the modding out kills all space-time supersymmetries, unless one of the
boosts 
happens to be  trivial.  This is the case if  $r_+=r_-$, which implies
$g_{\rm L} = {\bf 1}$.
The corresponding geometry describes the near-horizon region of 
an  extremal five-dimensional black hole \cite{mastro}.
Since the unbroken background supersymmetries are, however, in
this case  exclusively left-moving, any  D-brane will necessarily 
break  them all.
Indeed, the supercharges \eqref{unbr} cannot be defined
for any~$\Omega$.

This argument shows, nevertheless, how to construct supersymmetric D-branes
in a closely-related context.
 The idea is to replace the hyperbolic elements 
\eqref{hyper} by elliptic elements
 of $SL(2,\bR)$,  which act as real rotations
on the supercharges.
 Combining  these with equal-angle $SU(2)$ rotations,  
acting  on the  $S^3$ part of the space, leads to discrete isometries
that preserve both left- and right-moving supercharges.
A similar phenomenon is familiar from the study of 
supersymmetric branes at angles
\cite{angles}. 
Such  orbifolds of AdS$_3$$\times $$S^3$
 should  arise in the near-horizon geometry
of spinning five-dimensional black holes \cite{cl}. 
 The resulting CFT  could  admit appropriate supersymmetric D3-branes. 

\vskip 0.3cm
\noindent{\bf \large Aknowledgments}
\vskip 0.2cm
\noindent
We have benefited from
discussions with  Nicolas Couchoud, 
Mike Douglas,  Gary  Gibbons, 
Gary  Horowitz, Costas  Kounnas, Juan  Maldacena, Chris  Pope,  
Sylvain Ribault,  Steve Shenker, Sonia  Stanciu,  Arkady Tseytlin and
Paul Windey.  Research  supported
by the European networks ``Superstring theory"
(HPRN-CT-2000-00122), ``The quantum structure of space-time"
(HPRN-CT-2000-00131) and ``Physics across the present energy
frontier: probing the origin of mass" (HPRN-CT-2000-00148).


\vskip 1cm
\noindent{\bf \large Appendix A: Coordinate systems}
\vskip 0.5cm
\noindent
We have collected  in this appendix, 
for the reader's convenience,  the various  systems of coordinates
for AdS$_3$, which are  employed   in  the main text. 
When the same symbol is used in two different systems,  the
coordinate in question is the same. The coordinates $X^ M$ of
the embedding space are called, for short,  Cartesian coordinates.

\vskip 0.5cm
\noindent\underline{Cylindrical coordinates ($\tau,\rho,\phi$) } 
\vskip 0.2cm
\noindent 
These are good global coordinates of the entire covering space, 
with ranges
$\tau\in ]-\infty,+\infty[$, $\rho\in [0,\infty[$,
and $\phi\in [0,2\pi]$. They are related to the Cartesian
coordinates as follows:
$$
X^0 + i X^3 = L \cosh\rho \e^{i \tau} \ , \ \ X^1 + i X^2 =
 L \sinh\rho \e^{i\phi}\; . 
\eqno({\rm A.1})
$$
The metric in  cylindrical coordinates reads:
$$
ds^2 = L^2 \left(-  \cosh^2\! \rho \,
d\tau^2 + d\rho^2  + \sinh^2\! \rho\, d\phi^2 \right) .
\eqno({\rm A.2})
$$
The boundary of AdS$_3$  is at $\rho \to \infty$.
Its conformal
nature is more transparent in a system where the radial coordinate is
compact,
e.g. $\sinh \rho \equiv  \tan \vartheta$,  with $\vartheta \in
[0,\pi/2[$.
The conformal boundary is located in these coordinates at
$\vartheta \to \pi/2$,  and the metric takes  the following form:
$$
ds^2 = {L^2 \over \cos ^2\! \vartheta}
\left( - d\tau^2 +
d\vartheta^2 + \sin^2\! \vartheta \, d\phi^2 \right).
\eqno({\rm A.3})
$$
The hyperboloid  is
obtained by periodic identification of 
$\tau \cong \tau + 2\pi$.
The hyperbolic plane $H_3$ is obtained by the Wick rotation
$\tau \to  i\tau$.

\vskip 0.5cm
\noindent \underline{Poincar{\'e} coordinates ($t,x,u$) }
\vskip 0.2cm
\noindent  
These render explicit the two-dimensional Poincar{\'e}
invariance of the space.
They  are related to the Cartesian coordinates as follows:
$$
X^0+X^1 = Lu\ , \ \  X^2\pm X^3 = Luw^\pm \ ,  \ \ 
X^0-X^1 = L\left({1\over u} + u\,w^+w^-\right) , 
\eqno({\rm A.4})
$$
where $w^\pm \equiv  x\pm t$.
In terms of 
cylindrical coordinates they are given by
$$
w^\pm = {1\over u}\left(
\sinh \rho \, \sin \phi \pm \cosh \rho \, \sin \tau
\right)
\eqno({\rm A.5})
$$
and
$$
u = 
\cosh \rho  \, \cos \tau + \sinh \rho  \, \cos \phi  \, .
\eqno({\rm A.6})
$$
The metric in Poincar{\'e} coordinates has the standard form:
$$
ds^2 = L^2 \left(
 { du^2\over u^2} +  u^2 \,  dw^+ dw^- \right).
\eqno({\rm A.7})
$$
As ($t,x,u$)  range over the entire ${\bf R}^3$, they cover exactly
once the hyperboloid.
The  boundary of AdS$_3$ is at $\vert u\vert \to \infty$. The surface
$u=0$ is
a Rindler horizon.
The hyperbolic plane $H_3$ is obtained by Wick rotating $w^+$ and $w^-$ to
complex-conjugate coordinates $z$ and $\bar z$.  The upper
half space $u\geq 0$,  covers the entire hyperbolic plane.

\vskip 0.5cm
\noindent\underline{Anti-de Sitter coordinates ($\tau,\psi,\omega$) }
\vskip 0.2cm
\noindent 
These are such  that the fixed-$\psi$ slices are 
AdS$_2$ space-times. 
In terms of the Cartesian embedding coordinates they are defined by
$$
X^1 = L \sinh \psi \ , \ \ 
X^2 = L \cosh \psi \,\sinh \omega  \ , \ \ 
X^0 + i X^3 = L \cosh \psi \, \cosh  \omega \, \e^{i \t} \, .
\eqno({\rm A.8})
$$
The relation  to  cylindrical coordinates is
$$
\sinh \rho \, \cos \phi = \sinh \psi \ , \ \ 
\cosh \rho = \cosh \psi \, \cosh \omega \, .
\eqno({\rm A.9})
$$
The coordinates ($\tau,\psi,\omega$)   are  good  global coordinates
 that  take their values in the entire ${\bf R}^3$. 
The metric in this system reads:
$$
ds^2 = L^2  \left[ d\psi^2 + \cosh^2\! \psi \left(
-\cosh^2\! \omega \, d\t^2 + d\omega^2 
\right)\right] .
\eqno({\rm A.10})
$$

Defining $v \equiv \sinh \psi$ we can write the metric
equivalently as
$$
ds^2 = L^2  \left[ {dv^2\over 1 + v^2} + \left(1 + v^2\right) \left(
-\cosh^2\! \omega \, d\t^2 + d\omega^2 \right)\right]\, .
\eqno({\rm A.11})
$$
Notice that  Poincar{\'e} coordinates also give a natural slicing
of AdS$_3$ in terms of  (constant-$x$) AdS$_2$ space-times. The radius  of
these latter is fixed and  equal to $L$,
whereas the  constant-$\psi$
slices have continuously-varying radius, $\ell= L \cosh\psi$. 

\vskip 0.5cm
\noindent \underline{De Sitter coordinates ($\tilde t, \tilde u, \phi$) }
\vskip 0.2cm
\noindent 
These are related to the Cartesian coordinates as follows:
$$
X^0 = L \cosh \tilde\psi \ , \ \ 
X^3 = L \sinh \tilde \psi \, \sinh  {\tilde t} \ , \ \ 
X^1+iX^2 = L \sinh \tilde \psi \,  \cosh   {\tilde t}   \e^{i\phi} \, .
\eqno({\rm A.12})
$$
For $\tilde\psi \in [0 , \infty[$, $\tilde t\in ]-\infty, +\infty[$ and
$\phi\in[0,2\pi]$, the patch covers only a part of the hyperboloid: half
the outside region of the light cone.

The further change of coordinates $\tilde u \equiv \sinh\tilde\psi$ puts the
metric in the form
$$
ds^2 = L^2 \left[
{ d{\tilde u}^2\over 1+ {\tilde u}^2} +  {\tilde u}^2 ( -d{\tilde t}^2 + 
\cosh^2{\tilde t}\,  d\phi^2  ) \right] .
\eqno({\rm A.13})
$$
Fixed-$\tilde u$ slices are two-dimensional de Sitter space-times with
radii $\ell= L\tilde u$. 

\vskip 0.5cm
\noindent\underline{Hyperbolic  coordinates ($\tilde\tau,\chi,\phi $) }
\vskip 0.2cm
\noindent 
For completeness we  give also this  system of coordinates,  in
which the natural slicing is in terms of hyperbolic $H_2$ planes. 
Its relation to  Cartesian coordinates is as follows:
$$
X^0 = L \sin {\tilde\tau} \ , \ \ 
X^3=  L \cos {\tilde \tau}\, \cosh \chi \ , \ \ 
X^1+iX^2 = L \cos {\tilde \tau} \,\sinh  \chi \e^{i\phi}\, ,
\eqno({\rm A.14})
$$
with $ {\tilde\tau} \in [0, 2\pi]$ and $\chi \in [0, +\infty[$. 
The  metric is 
$$
ds^2 = L^2 \left[ - d{\tilde\tau}^2 + 
\cos^2 \! {\tilde\tau} \left( d\chi^2 +
\sinh^2\! \chi \,  d\phi^2 \right) \right] .
\eqno({\rm A.15})
$$
The above 
coordinate patch covers only a part of the hyperboloid.
The complementary part can be covered by 
two de Sitter coordinate  patches. 

\vskip 0.5cm
\noindent\underline{BTZ coordinates ($t^\prime,r,\varphi$) }  
\vskip 0.2cm
\noindent 
The metric in these coordinates is that of the BTZ black hole,
$$
ds^2 =L^2 \left[ 
-f^2(r) \, {dt^\prime}^2 + f^{-2}(r) \, dr^2 + r^2 \left(
d\varphi-{r_+ r_- \over   r^2}\, dt^\prime\right)^2\right] ,   
\eqno({\rm A.16})
$$
where
$$
f(r) ={1\over r}\sqrt{\left( r^2_{\vphantom +}-r^2_{+} \right)
\left(r^2_{\vphantom -}-r^2_{-}\right)} \, . 
\eqno({\rm A.17})
$$
The coordinates $t^\prime$ and $\varphi$ should not be confused with
the  $t$ and $\phi$ of other coordinate systems. In the special case 
$r_+=r_-=0$ the BTZ and Poincar{\'e} coordinates ($u>0$) coincide.
In the region outside
the outer horizon, $r>r_+$, 
we can relate the BTZ to the Cartesian coordinates
as follows:
$$
X^0 \pm X^1 = \pm L \left({r^2_{\phantom +}  - r^2_{+}
\over r^2_+ - r^2_-}\right)^{1/2}  \exp \pm \left( 
r_+ t^\prime  - r_- \varphi \right) \, ,
\eqno({\rm A.18})
$$
$$
X^2 \pm X^3 = \pm L \left({r^2_{\phantom -}  - r^2_{-}
\over r^2_+ - r^2_-}\right)^{1/2}  \exp \pm \left( 
r_+ \varphi  - r_- t^\prime \right)\, .
\eqno({\rm A.19})
$$
In the interior region where one or both of the square roots become
imaginary, these  must be replaced by  their absolute values, and one 
should drop  the corresponding $\pm $ sign. We have here  assumed that
$r_+>r_-\geq 0$;  in the extremal case $r_+=r_-$ the above
transformations degenerate. 

The BTZ identification $\varphi  \cong \varphi+2\pi$ corresponds to
modding out by a $SL(2,{\bf R}) \times SL(2,{\bf R})$ boost, with
parameters $r_+$ and $r_-$. The BTZ coordinates can be defined more
generally by choosing a  Killing vector $\xi$
cooresponding to a ``double boost", and then posing 
$$
\partial_{\varphi}\equiv \xi   \ \ {\rm and} \ \ 
r^2 \equiv (\xi, \xi) \, .
\eqno({\rm A.20})
$$
Regions of AdS$_3$ where $(\xi,\xi)$ is negative must be excised
before the BTZ identification, in order to avoid closed time-like
curves. The boundary of such regions is the black-hole singularity.
Notice that the BTZ coordinates do not cover entirely even the
remaining part of the hyperboloid, where $\xi$ is space-like.


\vskip 1cm
\boldmath
\noindent{\bf \large Appendix B: AdS$_2$ branes in global coordinates}
\unboldmath
\vskip 0.5cm
\noindent
In this appendix we revisit the AdS$_2$ solutions in global
coordinates. General static solutions are easy  to analyse  in
cylindrical coordinates, where they can be described by a function
$\rho(\phi)$ and a world-volume ``electric" field $F_{\t\phi}$.
The Gauss condition plus the continuity equation for the
energy--momentum tensor, $\Theta^\alpha_{\; \beta}$, give a simple
first-order differential equation,  which can be, in principle, solved.
The current of $\phi$-momentum, $\Theta^\phi_{\;  \phi}$, 
is a world-sheet constant in the static case.
Note that contrary to the current $\Theta_{\; x}^{x}$ in Poincar{\'e}
coordinates,  $\Theta^\phi_{\;  \phi}$  does not vanish
for the general AdS$_2$ solution: 
this is consistent with the fact that 
$\partial_\phi$ is not a direction
transverse to the D-string when $C\not=0$.

Though  the cylindrical coordinates are simple enough,  the global
anti-de Sitter system $(\tau,\psi,\omega)$  is even better
adapted for discussing the AdS$_2$ world-volumes. These 
 are spanned by the coordinates 
$(\t ,\omega)$  at some fixed  value
of $\psi$. We will now determine this constant value in terms of the number
$q$ of bound fundamental strings. In    
anti-de Sitter coordinates the
ambient   metric is given by  Eq. (A.10), while the
Neveu--Schwarz field strength background  reads:
$$
H  \equiv dB = 2L^2 \cosh^2\! \psi \,
\cosh \omega \, d\psi \wedge d\omega \wedge d\t \, .
\eqno({\rm B.1}) 
$$
We can choose
$$
B =   L^2 \left( {\sinh 2 \psi \over 2} +  \psi \right) 
\cosh \omega \,  d\omega  \wedge  d\t \, .
\eqno({\rm B.2}) 
$$
The D-string carries a (covariantly-constant) world-volume  electric
field, which in static coordinates takes the form  
$$
F = dA =  - \kappa \cosh \omega \, d\omega \wedge d\t \, ,
\eqno({\rm B.3})
$$
with $\kappa$ an a priori arbitrary constant. 
To fix $\psi$ as function of $\kappa$, consider the 
 Dirac--Born--Infeld action  of the D-string,  
$$
I(\kappa,\psi)  =  L^2 T_{\rm D} \int 
d\tau\,  d\omega\,
\cosh \omega  \, \sqrt{\cosh^4 \! \psi 
- \left( {\sinh 2 \psi\over 2} + \psi 
- {2\pi \alpha^\prime \kappa \over L^2} \right)^2}\, .
\eqno({\rm B.4})
$$
Notice that we are here working in a ``mini-superspace"  approximation --
fortunately, 
the  unbroken $SL(2,\bR)$ invariance  garantees that
if we extremize with respect to a constant
$\psi$ we will find  a solution to the full
equations of motion.  Performing the variation of (B.4) gives 
the extremum: 
$$
\psi_0 =  { 2\pi \alpha^\prime \kappa \over L^2}.
\eqno({\rm B.5})
$$
Comparing the AdS$_2$ radius,  $\ell=L\cosh\psi_0$,  with the results of
Section
4 leads to the relation
$$
\sinh\psi_0 = C = q{T_{\rm F}\over T_{\rm D}}\, .
\eqno({\rm B.6})
$$
Alternatively, this quantization of the AdS$_2$
world-volumes could have been derived directly from the
Gauss condition,  Eq. (4.12). 
Since  $T_{\rm D}\sim 1/\lambda_s$, the  quantization is invisible in 
the CFT at  disk level. 

The stability analysis of quadratic  fluctuations can be readily performed, 
along the lines of Ref. \cite{bds}, in the
anti-de Sitter  coordinate system. 


\vskip 1cm
\boldmath
\noindent{\bf \large Appendix C:  $H_2$ instantons }
\unboldmath
\vskip 0.5cm 
\noindent
For completeness, we describe here briefly
the instantonic $H_2$ solutions.
The analysis is simplest in hyperbolic coordinates, in which 
the  metric and antisymmetric tensor backgrounds can be written
$$
ds^2 = L^2 \left[ - d\tilde\tau^2 + 
\cos^2 \! \tilde\tau  \left( d\chi^2 +
\sinh^2\! \chi \,  d\phi^2 \right) \right]
\eqno({\rm C.1})
$$
and
$$
B =  L^2 
\left({\sin 2 \tilde\tau \over 2} + \tilde\tau\right) 
\sinh \chi \, d\phi \wedge d\chi \, . 
\eqno({\rm C.2})
$$
The Euclidean D-branes are  spanned by  $(\chi,\phi)$
at a constant value of the time coordinate $\tilde\tau$.
They carry a uniform  world-volume magnetic  flux
$$
F = dA = -\tilde\kappa \sinh \chi \,  d\phi  \wedge d\chi \, .
\eqno({\rm C.3})
$$
The DBI action is purely imaginary and infinite, 
$$
 I (\kappa,\tilde\tau) = 2\pi i  L^2 T_{(2)}
\sqrt{\cos^4 \! \tilde\tau + \left({\sin 2\tilde\tau \over 2} +\tilde\tau 
-{2 \pi \alpha^\prime\tilde\kappa \over  L^2}\right)^2}\,
 \int_0^{\infty} d\chi \sinh \chi \, .
\eqno({\rm C.4})
$$
Extremizing it formally  gives:
$$
\tilde\tau_0 = {2 \pi \alpha^\prime \tilde\kappa \over L^2}\, .
\eqno({\rm C.5})
$$
The unbroken $SL(2,\bR)$ invariance again garantees that this is an
exact saddle point of the action. 
After Wick rotation, both $\tilde\tau_0$ and $\tilde\kappa$ become
 pure imaginary.

The physical interpretation of these solutions is unclear. The uniform
world-volume flux  implies a uniform  density of D-instantons, which is
consistent with the divergence of the Euclidean action.

\end{document}